\documentclass[prb,twocolumn,aps,floats,superscriptaddress,amsmath,amssymb,showpacs]{revtex4}
\usepackage{graphicx}
\usepackage{epsfig}
\usepackage{epsf}

\begin{document}

\def\beq{\begin{equation}}
\def\eeq{\end{equation}}
\def\be{\begin{equation}}
\def\ee{\end{equation}}

\def\iomn{i\omega_n}
\def\iom#1{i\omega_{#1}}
\def\c#1#2#3{#1_{#2 #3}}
\def\cdag#1#2#3{#1_{#2 #3}^{+}}
\def\epsk{\epsilon_{{\bf k}}}
\def\Ga{\Gamma_{\alpha}}
\def\Seff{S_{eff}}
\def\dinf{$d\rightarrow\infty\,$}
\def\T{\mbox{Tr}}
\def\t{\mbox{tr}}
\def\cG0{{\cal G}_0}
\def\cS{{\cal S}}
\def\divnum{\frac{1}{N_s}}
\def\vac{|\mbox{vac}\rangle}
\def\intR{\int_{-\infty}^{+\infty}}
\def\intb{\int_{0}^{\beta}}
\def\spinup{\uparrow}
\def\spindown{\downarrow}
\def\bra{\langle}
\def\ket{\rangle}

\def\ka{{\bf k}}
\def\vk{{\bf k}}
\def\vq{{\bf q}}
\def\vQ{{\bf Q}}
\def\vr{{\bf r}}
\def\q{{\bf q}}
\def\R{{\bf R}}
\def\kp{\bbox{k'}}
\def\a{\alpha}
\def\b{\beta}
\def\d{\delta}
\def\D{\Delta}
\def\e{\varepsilon}
\def\ed{\epsilon_d}
\def\ef{\epsilon_f}
\def\g{\gamma}
\def\G{\Gamma}
\def\l{\lambda}
\def\L{\Lambda}
\def\o{\omega}
\def\ph{\varphi}
\def\s{\sigma}
\def\chib{\overline{\chi}}
\def\et{\widetilde{\epsilon}}
\def\hn{\hat{n}}
\def\hnu{\hat{n}_\uparrow}
\def\hnd{\hat{n}_\downarrow}

\def\hc{\mbox{h.c}}
\def\Im{\mbox{Im}}

\def\est{\varepsilon_F^*}
\def\v2o3{V$_2$O$_3$}
\def\uc2{$U_{c2}$}
\def\uc1{$U_{c1}$}

\def\bea{\begin{eqnarray}}
\def\eea{\end{eqnarray}}
\def\#{\!\!}
\def\@{\!\!\!\!}

\def\vi{{\bf i}}
\def\vj{{\bf j}}

\def\+{\dagger}


\def\up{\spinup}
\def\down{\spindown}


\def\e{\epsilon}

\def\veff{V_{\rm{eff}}}
\def\deff1{D_{1\rm{eff}}}
\def\d2eff{D_{2\rm{eff}}}

\title{Orbital selective Mott transition in multi-band systems:\\
slave-spin representation and dynamical mean-field theory}

\author{L.~de' Medici}
\affiliation{Centre de Physique Th{\'e}orique, {\'E}cole Polytechnique
91128 Palaiseau Cedex, France}
\affiliation{Laboratoire de Physique des Solides, CNRS-UMR 8502, Universit{\'e} de Paris-Sud, B{\^a}timent 510, 91405 Orsay, France}
\author{A.~Georges}
\affiliation{Centre de Physique Th{\'e}orique, {\'E}cole Polytechnique
91128 Palaiseau Cedex, France}
\author{S.~Biermann}
\affiliation{Centre de Physique Th{\'e}orique, {\'E}cole Polytechnique
91128 Palaiseau Cedex, France}

\begin{abstract}
We examine whether the Mott transition of a half-filled,
two-orbital Hubbard model with unequal bandwidths occurs
simultaneously for both bands or whether it is a two-stage process
in which the orbital with narrower bandwith localizes first
(giving rise to an intermediate `orbital-selective' Mott phase).
This question is addressed using both dynamical mean-field theory,
and a representation of fermion operators in terms of slave
quantum spins, followed by a mean-field approximation (similar in
spirit to a Gutzwiller approximation). In the latter approach, the
Mott transition is found to be orbital-selective for all values of
the Coulomb exchange (Hund) coupling $J$ when the bandwidth ratio
is small, and only beyond a critical value of $J$ when the
bandwidth ratio is larger. Dynamical mean-field theory partially
confirms these findings, but the intermediate phase at $J=0$ is
found to differ from a conventional Mott insulator, with spectral
weight extending down to arbitrary low energy. Finally, the
orbital-selective Mott phase is found, at zero-temperature, to be
unstable with respect to an inter-orbital hybridization $V$, and
replaced at small $V$ by a state with a large effective mass (and a low
quasiparticle coherence scale) for the narrower band.

\end{abstract}


\maketitle


\bigskip

\section{Introduction}

The Mott metal-insulator transition plays a central role in the
physics of all strongly correlated electron materials.
At a qualitative level,
localization of the electrons can occur when the kinetic energy
gain (typically given by the bare bandwidth) is smaller than the
cost in on-site repulsive Coulomb energy ($U$).
In recent years, dynamical mean-field theory
(DMFT)~\cite{georges_review_dmft}$^,$\cite{georges_strong,kotliar_dmft_physicstoday}
has provided
a consistent theoretical framework which has advanced our understanding
of this phenomenon~\cite{georges_review_dmft,georges_mott_iscom},
in particular through the study of simplified models such as the one-band Hubbard model.

In real materials however, such as transition metal oxides,
several orbital components are involved.
Crystal-field effects and the Coulomb exchange energy ($J$) affect
the energy of on-site atomic states, which no longer depend only on the
total local charge as in the orbitally degenerate case. Furthermore,
the inter-site hopping amplitudes can be significantly different for
different orbital components (due e.g to their relative orientations).
It is therefore essential to understand how
these effects can affect the Mott transition, and whether qualitatively new
effects are possible when the orbital degeneracy is lifted.

Recently, this question has attracted a lot of attention.
In their study of Ca$_{2-x}$Sr$_x$RuO$_4$,
Anisimov {\it et al.}~\cite{Anisimov_OSMT} suggested
that a partial localization could take place, in which some
orbital components (with broader bandwidth) are conducting, while others
(with narrower bandwidth) are localized (see also Ref.~\cite{terakura_ruthenates}).
Following this proposal, several
studies have been performed in the model context, with controversial
results~\cite{Liebsch_OSMT_0, Liebsch_OSMT, Liebsch_OSMT_2,Koga_OSMT,
Koga_OSMT_SCES}. Liebsch~\cite{Liebsch_OSMT_0,
Liebsch_OSMT} initially challenged the existence of such an
``orbital-selective Mott transition'' (OSMT), on the basis of DMFT
calculations.
Koga and coworkers~\cite{Koga_OSMT,
Koga_OSMT_SCES}, on the other hand, did find an OSMT within their DMFT calculations,
and suggested that a unique transition is found only if $J=0$. A symmetry argument was
put forward to explain this finding.

In this paper, a clarification of this problem is attempted, using DMFT and another,
complementary approach. The latter is based on a representation of fermion
operators in terms of slave quantum spins, specifically forged to address multi-orbital
models (Sec.~\ref{sec:slave_mft}). A mean-field approximation based on this representation,
similar in spirit to the Gutzwiller approximation, provides a fast and efficient method
in order to investigate the Mott transition in a wide range of
parameters (Sec.~\ref{sec:slave_osmt}). In Sec.\ref{DMFT}, a detailed
study of the previously unexplored regime in which one of the bands is much narrower
than the other and $J=0$ is presented, using exact diagonalizations and
quantum Monte Carlo methods in the DMFT framework.
Finally (Sec.\ref{sec:hyb}), the effect of an inter-band hybridisation
is considered.

\section{Model}

The model considered in this paper is a tight-binding model for two
bands, coupled by local interactions.
The Hamiltonian reads $H=H_0+H_{int}$, where $H_0$ is the non-interacting part:
\be\label{non-int_H}
H_0=-\sum_{m=1,2} t_m
\@ \sum_{<ij>,\s}d^\+_{im\s}d_{jm\s}+ {\rm h.c}
+\sum_{i,m\s}(\e_m-\mu) d^\+_{im\s}d_{im\s},
\ee
in which $d^\+_{im\s}$ ($d_{im\s}$) creates (annihilates) an electron
on the site $i$, in the orbital $m$, with spin $\s$.
The $\e_m$'s are crystal-field levels and $\mu$ is the chemical potential, kept here
for generality. In most of the paper however, we shall focus on the case of zero crystal-field
splitting ($\e_1=\e_2=0$) and half-filling of each band (i.e one electron
per site in each orbital, which corresponds to $\mu=0$ given our normalization of the
interaction term). At the end of the paper, we shall also consider the possibility
of a non-zero inter-orbital hybridization.

The full interaction, in the case of degenerate bands in a cubic
environment \cite{fresard_multiorbital_prb_1997,Castellani_V2O3}
reads:
\bea\label{H_int}
H_{int}\@&\,=\,\@&U \sum_{im}\tilde n_{im\up}\tilde n_{im\down}+U'\sum_{i\s}\tilde n_{i1\s}
\tilde n_{i2\bar\s}\nonumber \\
\@&+\@&(U'-J)\sum_{i\s}\tilde n_{i1\s} \tilde n_{i2\s}\nonumber \\
\@&-\@&J\sum_i\left[d^\+_{i1\up}d_{i1\down}d^\+_{i2\down}d_{i2\up} +
d^\+_{i1\down}d_{i1\up}d^\+_{i2\up}d_{i2\down}\right]\nonumber \\
\@&-\@&J\sum_i\left[d^\+_{i1\up}d^\+_{i1\down}d_{i2\up}d_{i2\down} +
d^\+_{i2\up}d^\+_{i2\down}d_{i1\up}d_{i1\down}\right],
\eea
where $\tilde n_{im\s}\equiv n_{im\s}-1/2$.
Following Castellani et al.~\cite{Castellani_V2O3},
the reduction of the Coulomb interorbital
Coulomb interactions $U^{\prime}$ as compared to the interorbital $U$
is related to the Hund's coupling $J$ by:
\be
U'=U-2J
\ee
In the case of vanishing Hund's rule coupling $J=0$ the interaction
vertex ($=U(n_1+n_2-2)^2/2$) thus depends only on the total charge, while if $J\neq 0$ the
inter-orbital interaction is weaker
than the intraorbital one and becomes sensitive to the
spin configuration.

\section{Slave-spin mean-field theory}
\label{sec:slave_mft}

\subsection{Slave-spin representation}

In this section, we introduce a new representation of fermion operators in terms
of constrained (``slave'') auxiliary fields, which proves to be particularly convenient
in order to study the multi-orbital hamiltonian above.
The main idea at the root of any slave-variable representation is to enlarge the
Hilbert space and to impose a local constraint in order to eliminate the unphysical
states. When the constraint is treated on average, a mean-field approximation is obtained.
Different slave-field representations will lead to different mean-field theories.
The quality of the mean-field approximation can be improved by tailoring the choice of
slave fields to the specific problem under consideration.
In general, a compromise has to be found between the simplicity of the representation, the
number of unphysical states which are introduced and the possibility of an analytical treatment of
the resulting mean-field theory.

For finite-U Hubbard models, Kotliar and Ruckenstein
\cite{kotliar_ruckenstein} have introduced a slave-boson
representation which can be used in the present context, when appropriately generalised
to multi-orbital models (in the spirit of the Gutzwiller approximation
\cite{gebhard_gutzwiller}).
However, this method introduces many variational parameters.
On the opposite, S.Florens and one of the authors introduced a very economical
representation of the N-orbital Hubbard model with $SU(N)$ symmetry based on a
single slave variable, taken to be the phase conjugate to the total charge on a given
lattice site (slave rotor representation)
\cite{florens_rotors_imp_2002_prb,florens_rotor_long_2004} .
However, this method is not appropriate when
the orbital symmetry is broken, as in the present work.

Here, we introduce a new slave-variable representation~\footnote{Although
the particle-hole symmetry of the
present model gives rise to simplifications, e.g on the
implementation of the constraints ($\lambda_{im\s}=0$), we present
here the general procedure.}
especially suited for dealing with multiband models, and addressing orbital-dependent
properties.
%
%
The basic observation behind this scheme is that the two possible occupancies of a
spinless fermion on a given site,
$n_d=0$ and $n_d=1$, can be viewed
as the two possible states of a spin-1/2 variable, $S^z=-1/2$ and $S^z=+1/2$.
This representation has been widely used in the case of hard-core bosons.
In the fermionic context however, one needs to insure anticommutation properties, and this is
done by introducing an auxiliary fermion $f$, with the additional local constraint:
\be
f^\dagger f = S^z + \frac{1}{2}
\ee
In this manner, one obtains a faithful representation of the Hilbert space, which reads:
\begin{eqnarray}
&|0\rangle = |n_f=0,S^z=-1/2\rangle \\
&|1\rangle\equiv d^\dagger |0\rangle = |n_f=1,S^z=+1/2\rangle
\end{eqnarray}
This constraint eliminates the two unphysical states
$|n_f=0,S^z=+1/2\rangle$ and $|n_f=1,S^z=-1/2\rangle$.
This representation is easily extended to the multi-orbital case,
by treating each orbital and spin species in this manner. Hence a set of
$2N$ spin-1/2 variables $S^z_{m\sigma}$ and auxiliary fermions
$f_{m\sigma}$ are introduced ($m=1,\cdots,N$ is the number of orbitals), obeying the
local constraint on each site:
\be\label{eq:constraint}
\hat{n}^f_{im\s}=S^z_{im\s}+\frac{1}{2},
\ee
This constraint can e.g be imposed with Lagrange multipliers fields
$\lambda_{im\sigma}(\tau)$.

We now explain how to rewrite the original hamiltonian (\ref{non-int_H},\ref{H_int}) in terms of
the slave spins and auxiliary fermions. We consider first for simplicity the case $J=0$, since
the case $J\neq 0$ requires an additional approximation, as discussed later.
For $J=0$, the interaction involves only the total electron charge on a given site, and
therefore reads:
\be\label{H_int_Spin_J=0}
H_{int}^{J=0}\equiv \frac{U}{2}\sum_i \left(\sum_{m,\s}\tilde{n}_{im\s}\right)^2
=\frac{U}{2}\sum_{i}\left(\sum_{m,\s} S^z_{im\s}\right)^2
\ee
In order to express the non-interacting part of the hamiltonian, we need to choose an
appropriate representation of the creation operator of a physical electron, $d^\dagger_{im\s}$.
There is some freedom associated with this, since different operators in the enlarged
Hilbert space spanned by the slave-spin and auxiliary fermions can have the
same action on the physical (constrained) Hilbert space.
We have not used the obvious possibility $d^\dagger\rightarrow S^+f^\dagger$,
$d\rightarrow S^{-}f$. This representation is correct in the physical Hilbert space (i.e when the
constraint is treated exactly), but it can be shown that additional mean-field approximations
based on this representation will ultimately lead to a problem with spectral weight conservation
because $S^+$ and $S^{-}$ do not commute.
Instead, we have chosen the representation
$d^\dagger\rightarrow 2S^x f^\dagger$, $d\rightarrow 2S^x f$, which is identical to the previous
one on the physical Hilbert space, and involves commuting slave-spin operators.
With this choice, the non-interacting part of the hamiltonian reads:
\bea
H_0\,\@&=\@&\,-\sum_m t_m \@ \sum_{<ij>,\s} 4S^x_{im\s}S^x_{jm\s}(f^\+_{im\s}f_{jm\s}+ h.c)\nonumber \\
\@&+\@&\sum_{i,m\s}(\e_m-\mu) f^\+_{im\s}f_{im\s}\nonumber
\eea
At this stage, no approximation has been made, provided the constraint is treated exactly.

\subsection{Mean-field approximation}

Approximations will now be introduced, which consists in three main steps:
i) treating the constraint on average, using a static and site-independent Lagrange
multiplier $\lambda_{m\s}$ ii) decoupling the auxiliary fermions and slave-spin degrees of
freedom, and finally iii) treating the slave-spin hamiltonian in a single-site mean-field
approach. This last step is quite independent of the two previous ones, and can be
rather easily improved on, as done in \cite{florens_rotor_long_2004}.

After the first two steps, one obtains two effective hamiltonians:
\bea\label{eq:H_f}
H_{eff}^f\,\@&=\@&-\sum_m t_m^{eff} \@ \sum_{<ij>,\s}(f^\+_{im\s}f_{jm\s}+ h.c.)\nonumber\\
\@&+\@&\sum_{i,m\s}(\e_m-\mu-\lambda_{m}) f^\+_{im\s}f_{im\s} \\
\label{eq:H_S}
H^S_{eff}\,\@&=\@&-\sum_m 4J^{eff}_m \@ \sum_{<ij>,\s}S^x_{im\s}S^x_{jm\s}\nonumber\\
\@&+\@&\sum_{i,m\s}\lambda_{m}(S^z_{im\s}+\frac{1}{2})+H_{int}[\{\vec{S}_{im\s}\}]
\eea
with $H_{int}[\{\vec{S}_{im\s}\}]=U/2 \sum_i(\sum_{im\s} S^z_{m\s})^2$ for $J=0$. In these expressions,
$t_m^{eff}$ and $J_m^{eff}$ are effective hoppings and slave-spin exchange constants which
are determined from the following self-consistency equations:
\bea\label{eq:scc}
&\@t_m^{eff} &\@= 4t_m\langle S^x_{im\s}S^x_{jm\s}\rangle \label{titilde}\\
&\@J^{eff}_m&\@= t_m\langle f^\+_{im\s}f_{jm\s}+f^\+_{jm\s}f_{im\s}\rangle\label{eq:Jeff}
\eea
The free fermion hamiltonian (\ref{eq:H_f}) describes the quasiparticle degrees of freedom.
Their effective mass is set by the renormalisation of the hopping:
$t_m^{eff}/t_m=4\langle S^x_{im\s}S^x_{jm\s}\rangle$. The quasiparticle weight is associated with a different
quantity, namely:
\be\label{eq:Zm}
Z_m = 4\langle S^x_{im\s}\rangle^2
\ee
Note that it depends in general on the orbital, a key feature for the physics that we want to
address with this technique.
Both the renormalisation of the mass and the quasiparticle weight are
self-consistently determined from the
solution of the quantum-spin hamiltonian (\ref{eq:H_S}), which describes the charge dynamics.
As clear from (\ref{eq:Zm}), metallic behaviour for orbital $m$ corresponds to long-range order in
$S^x_m$, while Mott insulating behaviour corresponds to $\langle S^x_m\rangle = 0$.

At this stage, the slave-spin degrees of freedom are still described by a quantum spin
hamiltonian on the lattice, and we therefore make the additional approximation (iii) of treating this
model on the level of a single-site mean-field.
We thus have to solve-the single-site spin hamiltonian:
\be\label{eq:Hs_mft}
H_s = \sum_{m\s}2h_m S^x_{m\s}+\sum_{m\s}\lambda_m(S^z_{m\s}+\frac{1}{2})+H_{int}[\vec{S}_{m\s}]
\ee
in which the mean-field $h_m$ is determined self-consistently from:
\be
h_m = 2z J_m^{eff} \langle S^x_{m\s} \rangle
\ee
where $z$ is the coordination number of the lattice.
This equation can be combined with (\ref{eq:Jeff}) to yield:
\be\label{eq:hm_scc}
h_m = 4 \langle S^x_{m\s} \rangle \frac{1}{{\cal N}}\sum_{\vk}\e_{\vk m}\langle f^\+_{\vk m\s}f_{\vk m\s}\rangle
\ee
In this expression, the fermionic expectation value is to be calculated with
the quasiparticle hamiltonian (\ref{eq:H_f}). Within this single-site mean-field
approximation however, the renormalisation of the hopping becomes identical to the
quasiparticle residue since $\langle S^x_{im\s}S^x_{jm\s}\rangle$ factorizes into
$\langle S^x_{im\s}\rangle^2$. As a result, the quasiparticle hamiltonian reads:
\be\label{eq:Hf_mft}
H^{f}_{eff} =
\sum_{\vk,m\s} (Z_m\e_{\vk m}+\e_m-\mu-\lambda_m) f^\+_{\vk m\s}f_{\vk m\s}
\ee
with $\e_{\vk m}\equiv -t_m /z \sum_{j, n.n(i)}e^{-\vk \cdot (\bf{i}-\bf{j})}$ the
Fourier transform of the hopping. Equations (\ref{eq:Zm},\ref{eq:Hs_mft},\ref{eq:hm_scc},\ref{eq:Hf_mft})
and the constraint equation (\ref{eq:constraint}) self-consistently determine the variational parameters
$h_m,\lambda_m$ and $Z_m=4 \langle S^x_{m\s}\rangle^2$. They are the basic mean-field equations
based on the slave-spin representation, which will be used below. Solving these equations requires
to diagonalize the single-site spin hamiltonian (\ref{eq:Hs_mft}), corresponding to a
$4^N\times 4^N$ matrix.

Let us finally discuss the case of a non-zero Hund's coupling $J\neq 0$. The first
three terms in (\ref{H_int}) are easy to treat in the slave-spin formalism, since they
involve only density-density interactions and are thus directly expressed in terms of the
Ising components of the slave spins. They read:
\be
\frac{U'}{2}\sum_{i}(\sum_{m,\s} S^z_{im\s})^2
+J\sum_{i,m}(\sum_{\s} S^z_{im\s})^2
-\frac{J}{2} \sum_{i,\s}(\sum_{m} S^z_{im\s})^2
\ee
In contrast, the ``spin-flip'' and (intra-site) ``pair-hopping''
terms (last two terms in (\ref{H_int})) are more difficult to deal with,
since they involve both slave-spin and auxiliary fermions operators. As a
result, four-fermion terms are introduced which require additional mean-field
decouplings. For simplicity, we choose to mimic the effect of these terms by
replacing them by operators which have exactly the same effect on the slave-spin
quantum numbers of the Hilbert space, namely:
\bea
\@&-\@&J\sum_i\left[S^+_{i1\up}S^-_{i1\down}S^+_{i2\down}S^-_{i2\up} +
S^+_{i1\down}S^-_{i1\up}S^+_{i2\up}S^-_{i2\down}\right]\nonumber \\
\@&-\@&J\sum_i\left[S^+_{i1\up}S^+_{i1\down}S^-_{i2\up}S^-_{i2\down} +
S^+_{i2\up}S^+_{i2\down}S^-_{i1\up}S^-_{i1\down}\right],
\eea
Despite the fact that these terms connect the physical and
unphysical parts of the Hilbert space (and therefore would strictly vanish
if the constraint was implemented exactly), it is reasonable to expect that they will
qualitatively describe the physics of the spin-flip and pair-hopping terms when
the constraint is treated on average, because their action on the slave-spin
quantum numbers is the correct one.
Hence, we shall use the following representation of the interacting part of the
hamiltonian for $J\neq 0$:
\bea\label{H_int_Spin}
H_{int}\@&\approx\@&\frac{U'}{2}\sum_{i}(\sum_{m,\s} S^z_{im\s})^2 \nonumber \\
\@&+\@& J\sum_{i,m}(\sum_{\s} S^z_{im\s})^2
-\frac{J}{2} \sum_{i,\s}(\sum_{m} S^z_{im\s})^2 \nonumber \\
\@&-\@&J\sum_i\left[S^+_{i1\up}S^-_{i1\down}S^+_{i2\down}S^-_{i2\up} + S^+_{i1\down}S^-_{i1\up}S^+_{i2\up}S^-_{i2\down}\right]\nonumber \\
\@&-\@&J\sum_i\left[S^+_{i1\up}S^+_{i1\down}S^-_{i2\up}S^-_{i2\down} + S^+_{i2\up}S^+_{i2\down}S^-_{i1\up}S^-_{i1\down}\right],
\eea

\subsection{Benchmarks}

In this section, we perform some benchmarks of the slave-spin representation and mean-field theory.

\subsubsection{Atomic limit (J=0)}

\begin{figure}[h]
\includegraphics[width=8cm]{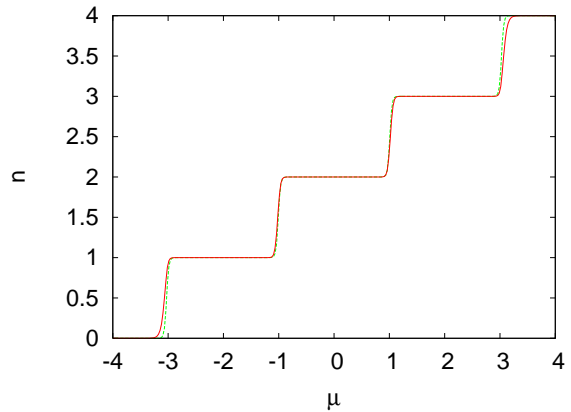}
\caption{\label{fig:Stair} (color online). Filling vs chemical
potential for a two-orbitals impurity (atomic limit of a
particle-hole symmetric Hubbard model), $U=2$, $\beta =50$: within
the slave spin mean field (full line) and exact result (dashed
line). The Coulomb staircase is correctly reproduced up to
temperatures of order $\sim U$}.
\end{figure}

In the $J=0$ case we check explicitly that the atomic limit
(i.e. $t_m=0$) of the degenerate N-band model with $SU(2N)$ symmetry is correctly reproduced.
Indeed our equations simplify drastically in this limit ($\tilde t_m=h_m=0$), leaving only
the $\lambda_{m\s}=\bar\lambda$ to be determined.
The constraint equation (\ref{eq:constraint}) reads in this case:
\be
n_F(\mu-\bar\lambda)=\mathcal{Z}^{-1}\sum^{2N}_{Q=0} \mathcal{N}_Q Q
e^{-\beta\left[\frac{U}{2}(Q-N)^2+\bar\lambda Q\right]}
\ee
where $n_F(\e)$ is the Fermi function,
$\mathcal{Z} \equiv \sum^{2N}_{Q=0}
\mathcal{N}_Q \exp{-\beta\left[\frac{U}{2}(Q-N)^2+\bar\lambda Q\right]}$,
$\mathcal{N}_Q\equiv \binom{2N}{Q}$ and Q is the total number of particles.
Solving  numerically this equation for $\bar\lambda$ leads to the
 correct ``Coulomb staircase'', as shown in Fig.\ref{fig:Stair}, as long
 as $T\ll U$.
At high temperatures the fact that we have imposed the constraints
only in average limits the accuracy, but in practice
$T\ll U$ is not a severe limitation.

\subsubsection{$N$-orbital Hubbard model with $SU(2N)$ symmetry and large-N limit}

\begin{figure}[h]
\includegraphics[width=8cm]{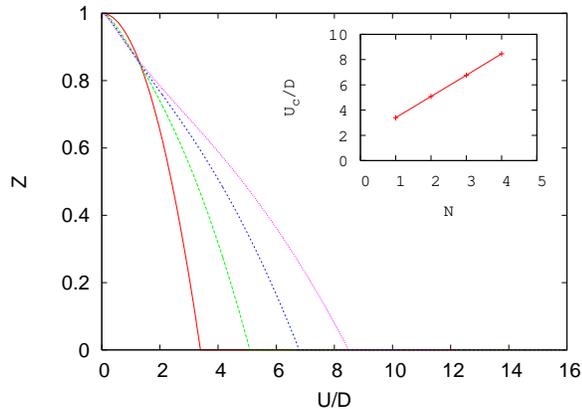}
\caption{\label{fig:Uc_vs_N} (color online). Quasiparticle weight,
obtained from slave-spin mean-field for the N-orbital Hubbard
model at half-filling (with, from left to right: $N=1,2,3,4$). The
non-interacting density of states is a semi-circle with half
bandwidth $D$ Inset: Dependence of the critical $U$ on $N$. The
exact large- N behaviour is obtained.}
\end{figure}

Here, we apply the slave-spin mean-field approximation to the N-orbital
model ($m=1,\cdots,N$), in the case where all bands have the same hopping,
with $J=0$. The results for the quasi-particle weight as a function of $U$,
at half-filling, are displayed on Fig.~\ref{fig:Uc_vs_N}. A transition into a
Mott phase is found for $U>U_c(N)$.
The exact large N behaviour of $U_c(N)$ in the limit of infinite coordination (DMFT)
is known~\cite{florens_orbital_2002_prb} to be linear in $N$,
the slope being $U_c/N=8\bar|\e|$, where $\bar\e\equiv\int_{-\infty}^0d\e D(\e)\e$.
This slope is correctly reproduced by the slave-spin mean-field approximation, indicating that
this approximation becomes more accurate as $N$ is increased.

One can actually calculate analytically the critical value of the coupling within this
approach, for arbitrary $N$, by performing a perturbative expansion around the atomic limit for small $h_m$.
This yields:
\be\label{eq:Uc_GA}
U_c=8(N+1)|\bar\e|
\ee
which coincides with the numerical determination in the inset of Fig.~\ref{fig:Uc_vs_N}.
We also note that (\ref{eq:Uc_GA}) is precisely the result of the Gutzwiller (slave-boson) approximation
in the multi-orbital case.

\subsection{Comparison with slave bosons and slave rotors}
\label{SB}

\begin{figure}[h]
\includegraphics[width=8cm]{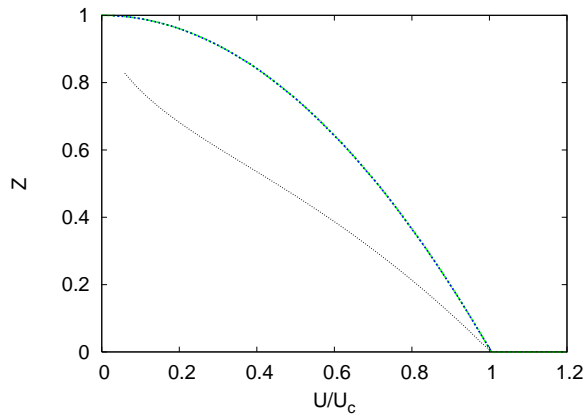}
\caption{\label{fig:Rotors_vs_Spins}  (color online).
Quasiparticle weight of the one-band Hubbard model, obtained with
slave rotors (thin line), slave spins, and the Gutzwiller (slave
boson) approximations (thick lines). The latter two actually
coincide. The small-$U$ behaviour of the slave-rotor approache is
due to the larger number of unphysical states (see text). }
\end{figure}

As suggested by the fact they yield identical values of $U_c$, the slave-spin
mean-field theory has many similarities with the Gutzwiller approximation (GA).
In fact, as shown on Fig.~\ref{fig:Rotors_vs_Spins}, the whole dependence of
$Z$ on $U$ is identical to that of the GA.

The slave spin representation has several advantages over the slave-boson
representations that can be used to formulate the GA. One advantage is
that the number of variables is smaller: $2N$ spin-$1/2$ degrees of freedoms
instead of $2^N$ slave bosons (one associated to each state in the Hilbert space,
in the absence of symmetries). Another advantage is that the number of unphysical
states is smaller than in slave boson representations, because the Hilbert space
spanned by the $2N$ quantum spins is finite by construction, while the Hilbert space
associated with the slave bosons in the absence of the constraint is
an infinite-dimensional one. This might be useful in considering finite-temperature
properties and the entropy of the model.

A similar remark applies when comparing the present slave-spin representation to
the slave-rotor representation recently developed by S.~Florens and one of the
authors
\cite{florens_rotors_imp_2002_prb,florens_rotor_long_2004} .
This representation is specifically tailored to $SU(2N)$ symmetric models,
and is very economical since it introduces only {\it one} slave variable.
Specifically, a (slave) quantum rotor and auxiliary fermions are introduced on each
site such that:
\be
d^\dagger_{im\s}=f^\dagger_{im\s}e^{i\theta_i}\,\,\,,\,\,\,
d_{im\s}=f_{im\s}e^{-i\theta_i}
\ee
where the phase is conjugate to
the local charge, corresponding to the local constraint:
\be
\sum_{m\s}(f^\dagger_{im\s}f_{im\s}-\frac{1}{2})=\hat{L_i}
\ee
in which $\hat{L_i}=1/i\partial/\partial\theta_i$ is the conjugate momentum to the
phase. It is clear from these expressions that there is a close similarity between
the slave-spin and slave-rotor formalisms. Two important differences must be noted:
(i) a single slave-rotor variable is introduced for all orbitals and (ii) the Hilbert space
of the unconstrained rotor is infinite-dimensional, containing an infinite tower
of charge states $|l\rangle$ which are physical only for $|l|\leq N$.
As a result, mean-field approximations in which the constraint is treated only on average
are less accurate when the contribution of these unphysical charge states become sizeable.
This is particularly true in the weak-coupling limit. On Fig.~\ref{fig:Rotors_vs_Spins},
we compare the slave-spin and slave-rotor result for the quasiparticle residue in the
one-band case, in order to illustrate this effect.

On the whole, slave-rotors and slave-spins offer two useful representations,
the former being very economical and well suited to situations in which only the
total local charge is involved (e.g in Coulomb blockade problems~\cite{florens_qdot_prb_2003}), while the
latter is well suited to the investigation of orbital-dependent properties, as in
the present article. Both methods are easy to implement at a very low numerical cost,
hence allowing for a fast and efficient investigation of the phase diagram and phase transitions
in a wide range of parameters.

\section{Orbital-selective Mott transition with slave-spin mean field theory}
\label{sec:slave_osmt}

In this section, we use slave-spin mean-field theory in order to
study the two-band model with unequal hoppings. The non-interacting density of
states of each band is
taken to be a semi-circle (of half-width $D_1=2t_1$ and $D_2=2t_2<D_1$),
corresponding to a Bethe lattice with infinite connectivity $z=\infty$
and nearest-neighbour hoppings $t_{1,2}/\sqrt{z}$.
No crystal-field splitting is introduced ($\e_1=\e_2=0$) and we
restrict ourselves to the case in which both bands are half-filled
($\langle n_1\rangle =\langle n_2\rangle=1$).
The model is thus particle-hole symmetric, implying that the chemical
potential $\mu=0$ and Lagrange multipliers $\lambda_{1}=\lambda_2=0$.
The parameter space was explored for $U>0$, $J=0\div 0.5U$ i.e. $U'=U-2J=0\div U$,
and the ratio between the two bandwidths $t_2/t_1=0\div 1.0$.
For the study of the Mott transitions in this model we monitor the quasiparticle weights
$Z_m=4\langle S^x_{m\s}\rangle^2$.

\begin{figure}[h]
\includegraphics[width=8cm]{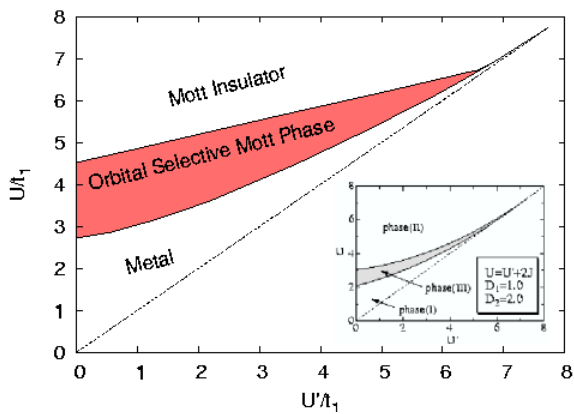}
\caption{\label{fig:Phase_diagram} (color online). Phase diagram
(U vs U') for $t_2/t_1=0.5$ at $T=0$ within the Slave Spins mean
field theory. Inset, same diagram obtained with Exact
diagonalization-Dynamical Mean Field Theory (ED-DMFT) in
\cite{Koga_OSMT}. The dotted line indicates $J=0$, i.e. $U=U'$. In
``phase I'' both bands are metallic, in ``phase II'' both bands
are insulating. ``Phase III'' is the orbital-selective Mott
phase.}
\end{figure}

Fig.~\ref{fig:Phase_diagram} displays the phase diagram within slave-spin mean-field theory
for the bandwidth ratio $t_2/t_1=0.5$.
Three different phases are found: at small U both bands are metallic (i.e. $Z_m\neq 0$),
at large U both are insulating ($Z_m=0$), and in between an ``orbital selective Mott phase'' (OSMP)
is found
in which only the band with largest bandwidth has $Z_1\neq 0$, while the narrower
band has $Z_2=0$.

In the inset of Fig.~\ref{fig:Phase_diagram}, we reproduce for comparison the result of
Koga {\it et al.}~\cite{Koga_OSMT} obtained within DMFT.
Qualitatively, one sees that the slave-spin mean-field compares rather well to
the DMFT results.
There are quantitative differences in the critical values of the couplings $U$ and
$U'$, a well known feature of Gutzwiller-like approximations. Also,
the linear dependence on $U'$ of the upper boundary of the OSMP phase is due to
the simplified treatment of the spin-flip and pair-hopping terms discussed above
(as indeed confirmed by the results of section \ref{Hund}).

There is however one significant qualitative difference between the slave-spin
results and those of Koga et al. (inset). We find that the endpoint of the
OSMP phase {\it does not lie exactly} on the $U=U'$ line.
Hence, within the slave-spin mean field, the Mott transition becomes orbital-selective
(OSMT) only when $J$ exceeds a critical value.
This is a rather significant finding, since
for $J=0$ the interacting part of the hamiltonian ($H_{int}$)
has full $SU(4)$ spin-orbital symmetry, while for
$J\neq 0$ the symmetry is lower. In
Ref.\cite{Koga_OSMT}, it was argued that indeed the enhanced symmetry of the $J=0$ case prevents
an orbital-selective Mott transition to occur.
Our finding that a critical value of $J$ is needed to induce an OSMT (for $t_2/t_1=0.5$)
suggests that symmetry considerations on $H_{int}$ may not be essential to the existence
of an orbital-selective transition. After all, the difference in bandwidths breaks the
$SU(4)$ symmetry from the kinetic energy part of the hamiltonian.
In order to study this issue in more detail, we perform in the next section a detailed
study of the $J=0$ case.

\subsection{OSMT at $J=0$}

In this section, we focus on the $J=0$ case, for which $H_{int}$ has full $SU(4)$ symmetry,
and explore the nature of the Mott transition in the full range of bandwidth ratio
from $t_2/t_1=0$ to $t_2/t_1=1$.

\begin{figure}[h]
\includegraphics[width=8cm]{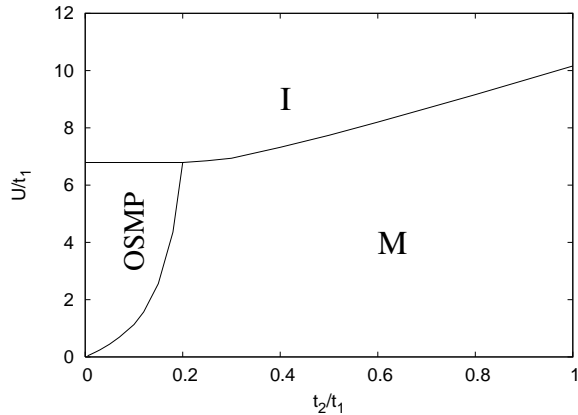}
\caption{\label{fig:Uc_vs_ratio} Dependence of the critical U on the ratio $t_2/t_1$ at $J=0$
and $T=0$. All transitions are second-order.}
\end{figure}

We find that the two bands undergo a common Mott transition at a single value of $U=U_c$
as long as the bandwidth ratio exceeds a critical threshold: $t_2/t_1>0.2$.
In contrast, for $t_2/t_1<0.2$, {\it an orbital-selective Mott phase is found}, despite the
enhanced symmetry of the interaction term.
Fig.~\ref{fig:Phase_diagram} displays our result for the phase diagram as a function of
$t_2/t_1$ and $U/t_1$. All transitions are found to be second-order when $J=0$.
On Fig.~\ref{fig:Uc_vs_ratio}, the quasiparticle weights
of each band is plotted as a function of $U$, for several values of $t_2/t_1$.
The localization of the narrower band manifests itself as a kink in the
quasiparticle weight of the wider band. As $U$ is increased further, the wide band
in turn undergoes a Mott transition.
We observe that, within slave-spin mean-field, the quasiparticle weight of the
wider band in the orbital-selective Mott phase coincides with that of a single-band model.
This is because the slave-spin mean-field neglects charge fluctuations of the localised orbital, so
that the physical behavior of the wide band becomes
effectively that of a one-band model as soon as the narrow band
becomes localized.
\begin{figure}
\includegraphics[width=8cm]{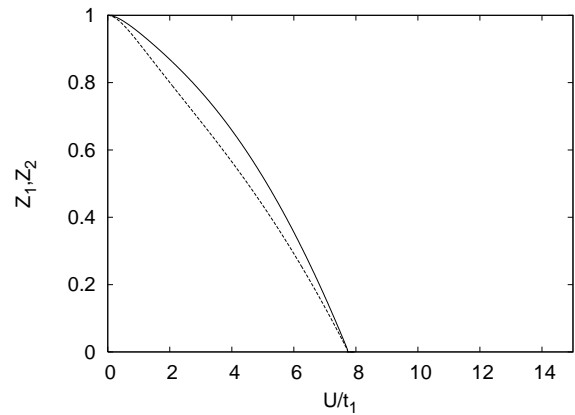}
\includegraphics[width=8cm]{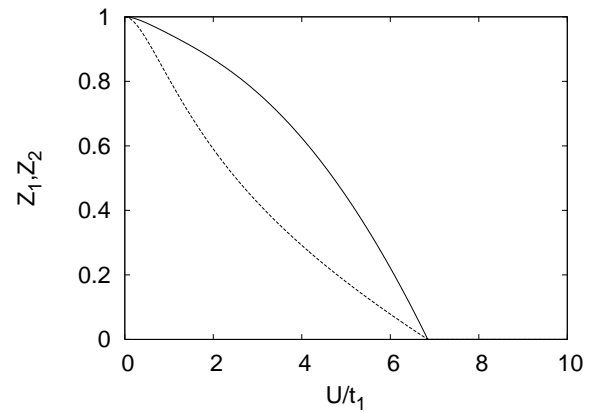}
\includegraphics[width=8cm]{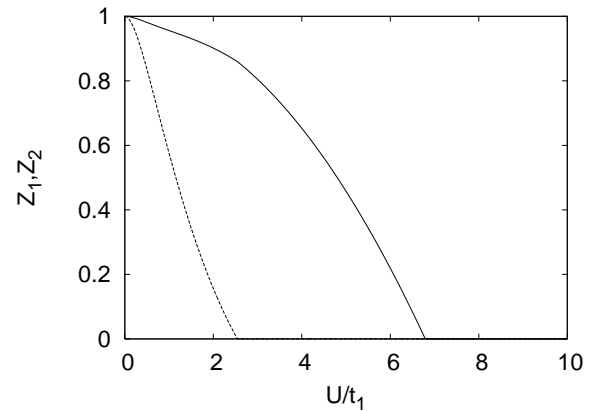}
\caption{\label{fig:Mott_MIT} $Z_m$ at $J=0$ and $T=0$ for $t_2/t_1=0.5$ (top), $0.25$ (middle), $0.15$ (bottom).}
\end{figure}

Our finding of an orbital-selective Mott transition at $J=0$ when $t_2/t_1$ is small enough,
within slave-spin mean-field theory, raises
two questions. First, is this finding an artefact of the slave-spin approximation or
does it survive a full DMFT treatment (i.e is it a genuine feature of the infinite-coordination
model ?). Second, does this invalidate the argument based on the symmetry of $H_{int}$ ?
The first question will be addressed in detail in Sec.~\ref{DMFT}, in which a DMFT study
will be performed, using exact diagonalization and Quantum Monte-Carlo techniques.
We will show that indeed, a transition does exist at $J=0$ when $t_2/t_1$ is small enough,
but that the nature of the intermediate phase (OSMP) at low-energy is a rather subtle issue.
In order to address the second question, let us briefly recall the symmetry argument of
Koga {\it et al.}~\cite{Koga_OSMT}. The argument relies on the gap to charge excitations in the insulating phase.
When $J=0$, charge excitations mix the two orbitals because of the enhanced $SU(4)$
symmetry. Instead, for $J\neq 0$, the charge excitations of lowest energy are independent
in each orbital sector. As a result, it is reasonable to expect (at least when the kinetic energy
term is treated in a perturbative manner) that the system can sustain
two different charge gaps when $J\neq 0$ while the gaps might coincide for $J=0$.
We observe however that this argument applies to the instability of the large-U Mott phase (in which
both bands are gapped) when $U$ is reduced, and suggests that, for $J=0$, the Mott gap closes at the same
value of $U$ for both bands when $U/t_1$ is reduced. It does not preclude however that a transition
into an intermediate phase does exist, in which the ``localised'' band (with the narrower bandwith)
is not fully gapped.
As we shall find below, there is
indeed clear evidence from the DMFT results that the orbital-selective Mott phase at $J=0$ is {\it not a
conventional Mott insulator} and that the narrow (``localised'') band does have spectral weight
down to zero-energy in this phase. Obviously, the slave-spin mean-field approach is too
rudimentary to be able to capture these fine low-energy aspects, but it is remarkable that
it does allow us to infer correctly that an intermediate phase is indeed present.

\subsection{Dependence of OSMT on J}

\begin{figure}[h]
\includegraphics[width=8cm]{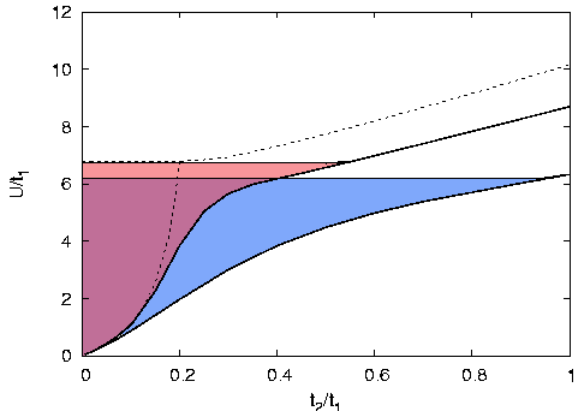}
\caption{\label{fig:UcJ_vs_ratio} (color online). Widening of the
OSMT zone with increasing $J/U$ at $T=0$. Transition lines are
shown for (from left to right) $J/U=0$ (dashed),$0.01,0.1$}
\end{figure}

\begin{figure}[h]
\includegraphics[width=8cm]{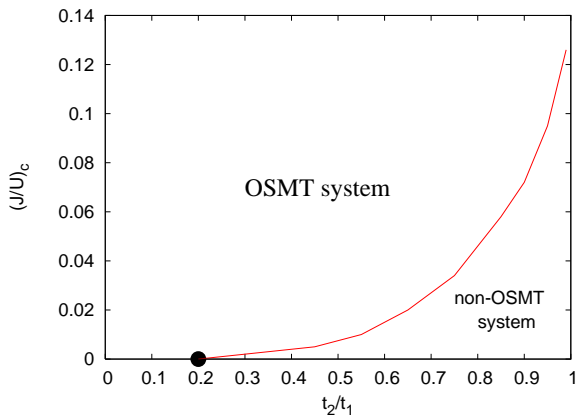}
\caption{\label{fig:Jc_vs_ratio} (color online). Dependence of the
critical $J/U$ above which the Mott transition becomes
orbital-selective, on the ratio $t_2/t_1$ at $T=0$.}
\end{figure}

Having clarified the situation for $J=0$, we come back to the effect of
a non-zero $J$, still within the slave-spin mean-field approximation.
Fig.~\ref{fig:UcJ_vs_ratio} shows how the phase diagram as a function of
the bandwidth ratio $t_2/t_1$ and of $U/t_1$ is modified for $J\neq 0$.
One sees that a finite $J$ {\it enlarges the orbital-selective Mott phase}
and favors an OSMT. The critical ratio $(t_2/t_1)_c$
below which an OSMP exists increases significantly, e.g.
$t_2/t_1\simeq 0.55$ for $J=0.01U$. With increasing $J$,
$(t_2/t_1)_c$ tends towards $1$. Hence a common Mott transition
for both bands is recovered for all values of $J$ only when $t_1=t_2$.
For a given bandwidth ratio $t_2/t_1$, the Mott transition is
orbital-selective for $J/U>(J/U)_c$. The dependence of this
critical ratio upon $t_2/t_1$ (i.e the location of the enpoint of the
OSMP phase) is displayed in Fig.~\ref{fig:Jc_vs_ratio}. (It should be noted however
that this critical ratio $(J/U)_c$ is underestimated by our simplified
treatment of the pair-hopping and spin-flip term). Finally, we found that
for finite $J$, the insulator to OSMP transition remains second-order, while the
metal to OSMP transition becomes first-order.

\subsection{Role of the spin-flip and pair-hopping
terms in the Hund Hamiltonian}\label{Hund}

\begin{figure}[h]
\includegraphics[width=8cm]{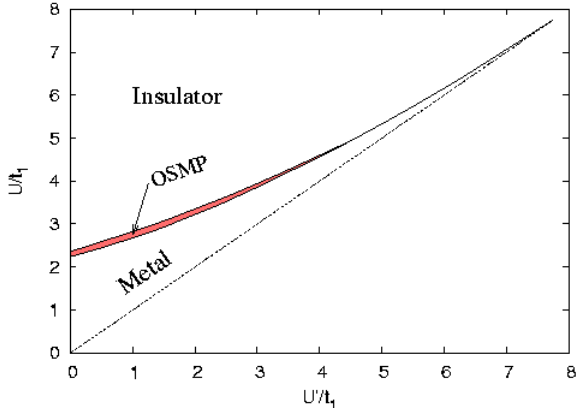}
\caption{\label{fig:Phase_diagram_Hund} (color online). Phase
diagram (U vs U') for
  $t_2/t_1=0.5$ at $T=0$ for a two-band Hubbard model without spin
flip and pair hopping terms in the interaction.}
\end{figure}

In order to clarify the role played by the different terms of the
interaction (\ref{H_int}), we have also
studied the Hamiltonian in which the spin-flip and the (on-site) inter-orbital
pair-hopping terms are dropped, namely:
\bea\label{H_int_Hund}
H_{int}\@&\,=\,\@&U \sum_{im}\tilde n_{im\up}\tilde n_{im\down}+(U-2J)\sum_{i\s}\tilde
n_{i1\s} \tilde n_{i2\bar\s}\nonumber \\
\@&+\@&(U-3J)\sum_{i\s}\tilde n_{i1\s} \tilde n_{i2\s}.
\eea
which is easily represented in terms of slave-spins as:
\bea
H_{int}\@&\,=\,\@&\frac{U'}{2}\sum_{i}(\sum_{m,\s} S^z_{im\s})^2+
\nonumber\\
\@&+\@& J\sum_{i,m}(\sum_{\s} S^z_{im\s})^2
-\frac{J}{2} \sum_{i,\s}(\sum_{m} S^z_{im\s})^2
\eea
This study is also motivated by a comparison to Quantum Monte-Carlo treatments
in which the spin-flip and pair-hopping terms are not easily treated.
On Fig.~\ref{fig:Phase_diagram_Hund} we display the slave-spin phase diagram
found for $t_2/t_1=0.5$ at zero-temperature. On sees that the OSMP shrinks
dramatically (albeit the two transitions do not actually merge).
This finding sheds light on the
results by Liebsch in ref.
\cite{Liebsch_OSMT_0,Liebsch_OSMT,Liebsch_OSMT_2}.
Because this study was based on Quantum-Monte Carlo,
hence neglecting the spin-flip and pair-hopping terms, it is natural that the orbital-selective
phase can be found only in a very narrow range of parameter space.
The key role of spin-flip and inter-orbital pair-hopping terms for the OSMT was
actually pointed out in the recent work of Koga et al. \cite{Koga_OSMT_SCES}.

\begin{figure}[h]
\includegraphics[width=8cm]{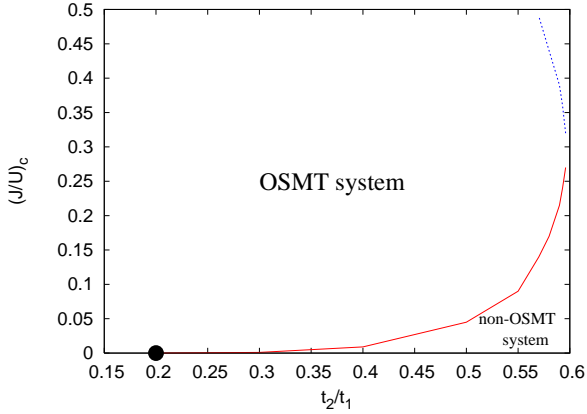}
\caption{\label{fig:Jc} (color online). Critical $J/U$ for the
model without spin-flip and pair-hopping as a function of
$t_2/t_1$. At small ratios of the bandwidths, the system displays
an OSMT above a critical $J/U$ ratio which vanishes at
$t_2/t_1=0.2$. In less anisotropic systems large Hund's coupling
are needed to realize OSMPs, whereas beyond the critical bandwidth
ratio of 0.6 no OMST is possible within this model.}
\end{figure}

In Fig.~\ref{fig:Jc}, we display the region in $t_2/t_1,J/U$ parameter space where the
Mott transition is found to be orbital-selective, analogously to Fig.\ref{fig:Jc_vs_ratio}.
The critical ratio $(J/U)_c$ is found to be roughly exponential in $t_2/t_1$.
In contrast to the case of the full hamiltonian (Fig.\ref{fig:Jc_vs_ratio}),
we find that no OSMT exists when $t_2/t_1$ exceeds a critical bandwidth ratio
$t_2/t_1\simeq 0.6$, for any $J/U$.
Beyond this value no OSMT is found at any $J/U$.
(Note that the upper critical line at large $J/U$ corresponds
however to the unphysical case of an attractive Coulomb interaction
due to $U-3J<0$).

\begin{figure}
\includegraphics[width=8cm]{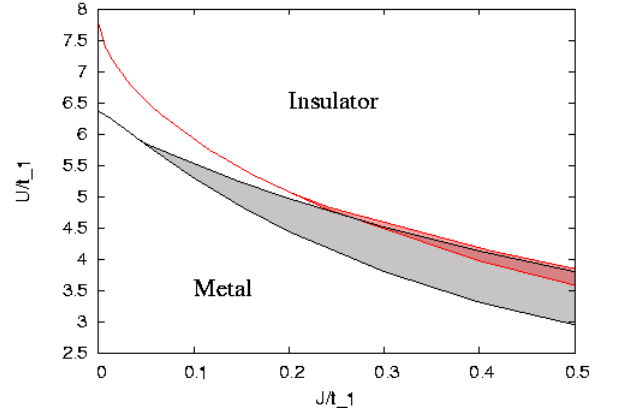}
\caption{\label{fig:Phase_diagram2} (color online). Same phase
diagram as in fig. \ref{fig:Phase_diagram_Hund} but in the U-J
plane at $T=0$ and at $\beta t_1=40$.}
\end{figure}

Finally, we emphasize that the orbital-selective Mott phase, which exists
only in a very narrow range of couplings for the simplified interaction (\ref{H_int_Hund})
at $T=0$, is actually enlarged at finite-temperature, as shown in
Fig.~\ref{fig:Phase_diagram2}.

\section{Dynamical- Mean Field Theory for $J=0$ and nature
of the OSMT phase}\label{DMFT}

In this section, we study the two-band model with $J=0$ using dynamical
mean-field theory. Our goal is to determine whether the transition into an
orbital-selective Mott phase (OSMP) found within the slave-spin approximation
at small enough $t_2/t_1$ is indeed a robust feature, and to shed light on
the possible low-energy physics of this phase.
Within DMFT, the lattice model is mapped onto a self-consistent two-orbital
Anderson impurity model~\cite{georges_kotliar_dmft,georges_review_dmft}
with effective action:
\bea
&-\int_0^\beta\int_0^\beta d\tau d\tau' \sum_{m\s}
d_{m\s}^\dagger(\tau){\cal G}_m^{-1}(\tau-\tau')d_{m\s}(\tau')+
\nonumber \\
&+\frac{U}{2}\int_0^\beta d\tau (n_1+n_2-2)^2
\eea
The hybridisations to the effective conduction bath are self-consistently
related to the local interacting Green's functions $G_m$ through:
\be
\label{eq:scc_dmft}
{\cal G}_m(\iomn)^{-1} = \iomn-t_m^2\, G_m(\iomn)
\ee
These equations are exact for an infinite-connectivity Bethe lattice
(corresponding to semi-circular non-interacting d.o.s). Particle-hole symmetry
with one electron per site in each band has been assumed.
We focus here on the paramagnetic solutions only.
The DMFT equations will be solved in the following using both
an exact diagonalisation (ED) and Quantum Monte-Carlo (QMC) technique.

\subsection{Exact diagonalisation study}
\label{sec:ED}

Within the adaptative exact-diagonalization
method~\cite{caffarel_ed_prl_1994,georges_review_dmft}, the effective
conduction-electron bath is discretized using a finite number of
orbitals $N_s$. Hence, one considers the two-orbital Anderson
impurity hamiltonian: \bea\label{Ham_AIM}
H_{AIM}\@&=&\@\sum_{m\s}\sum_{l=1}^{N_s} \e_{lm} a^\+_{lm\s}
a_{lm\s} +
\sum_{m\s}\sum_{l=1}^{N_s} V_{lm} (d^\+_{m\s} a_{lm\s} + h.c) \nonumber\\
\@&+&\@ \frac{U}{2} \left(\hat{n}_1+\hat{n}_2-2\right)^2
\eea
The operators $a_{lm\s},a^\+_{lm\s}$ describe the discretized conduction-
bath degrees of freedom.
The effective parameters $\{\e_{lm},V_{lm}\}$ have to be determined self-consistently,
according to (\ref{eq:scc_dmft}), namely:
\be\label{selfcons}
\sum^{N_s}_{l=1} \frac{\vert V_{lm}\vert ^2}{\iomn-\e_{lm}}
=t_m^2G_m(\iomn)
\ee
The ED method becomes an asymptotically exact solver of the DMFT equations in
the limit $N_s\rightarrow\infty$. In practice however, one can handle only a finite number of effective
sites. For the case at hand, we used a $T=0$ Lanczos algorithm, with $N_s=5$ (i.e 5 effective
sites per orbital). The self-consistency (\ref{selfcons}) is implemented on a
Matsubara grid corresponding to a (fictitious) inverse temperature $\beta$ (taken to
be in practice in the range $200$ to $500$, which insures a good resolution on the
low-energy physics).
We monitor in particular the quasiparticle weights, approximated as:
$Z_m= [1-\Im{\Sigma_m(i\omega_0)}/\omega_0]^{-1}$ (where $\omega_n=\pi/\beta(2n+1)$).

\begin{figure}[h]
\includegraphics[width=8cm]{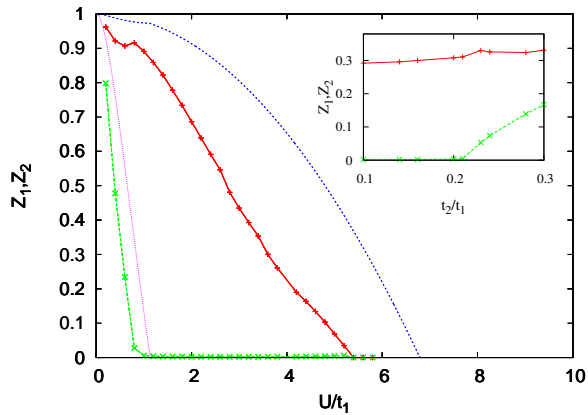}
\caption{\label{fig:Z1Z2_DMFT} (color online). Quasiparticle
residues at $J=0$ in DMFT(ED) for $t_2/t_1=0.1$ (symbols). For
comparison, the slave-spin mean-field approximation is also
displayed (continuous line). Inset: same quantities as a function
of $t_2/t_1$ at fixed $U/t_1=4.0$. (The kinks in $Z_1$ are
associated with the criticality of $Z_2$).}
\end{figure}

Our ED results for these quantities are displayed on Fig.~\ref{fig:Z1Z2_DMFT}, and
compared to the slave-spin results, both
as a function of $U$ for fixed $t_2/t_1$ and as a function of $t_2/t_1$
for fixed $U$ (inset).
It is clear from this figure that, within the energy resolution which
can be reached with $N_s=5$, an orbital-selective transition is indeed observed
in the DMFT(ED) results when $t_2/t_1$ is smaller than a critical value
(close to $0.25$), in remarkable agreement with
the slave-spin mean-field. The quantitative value of the critical coupling $U/t_1$
for the localisation of the wider band is overestimated, as usual in Gutzwiller-like schemes.

\begin{figure}[h]
\includegraphics[width=8cm]{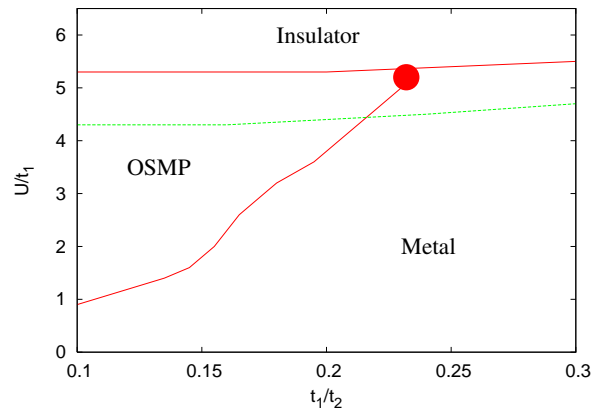}
\caption{\label{fig:DMFT} (color online). Phase diagram at $J=0$
and $T=0$ in
  ED-DMFT. The dashed line marks the disappearance of the Mott gap
  (downward runs).
For U values around 5 there is coexistence between an OSMP and -
depending on the bandwidth ratio - an insulating or metallic
phase.}
\end{figure}

The phase diagram obtained with ED, as a function of $t_1/t_2$ and $U/t_1$ is
displayed in Fig.~\ref{fig:DMFT}. Good qualitative agreement with the
slave-spin mean field is found (Fig.~\ref{fig:Uc_vs_ratio}).
We also studied the hysteresis properties by performing
runs for increasing and decreasing values of $U/t_1$.
This results in two almost parallel transition lines on Fig.~\ref{fig:DMFT},
one corresponding to the disappearence of the metallic solution
(obtained from a series of runs for increasing U), the other corresponding to the loss of the
insulating nature of the wide band (from a series of runs for decreasing U).
In the region between these two lines, coexistence of two types of DMFT solutions is found:
for small $t_2/t_1$, one of the solutions is orbital-selective Mott and the other is fully
insulating, while for larger $t_2/t_1$, one of the solutions is metallic and the other fully insulating.
The actual thermodynamic transition is given by the
crossing of the free energies of the two solutions.
In contrast, no hysteresis has been found at the transition between the
OSMP and metallic phases.
This suggests that the transition from the metallic to the OSMP phase is
of a very different nature than the Mott transition of the wider band.
If only the gap closure is monitored, then only one transition is found at $J=0$, in
agreement with the symmetry argument of Ref.~\cite{Koga_OSMT}. As we shall demonstrate below,
there is indeed
strong evidence that the orbital-selective Mott phase at $J=0$ {\it does not display a
sharp gap in either orbitals}.

\subsection{Low-energy nature of the orbital-selective Mott phase at $J=0$,
from ED and QMC}

In order to understand better the nature of the orbital-selective
phase found at $J=0$, we take a closer look at the local Green's
function for each orbital, in each phase. In order to do this, we
also solved the DMFT equations using the Quantum Monte-Carlo algorithm
of Hirsch and Fye \cite{hirsch_fye}. The ED and
QMC methods are quite complementary. The former applies at $T=0$ but suffers from a limited
energy- resolution due to the small value of $N_s$, while the latter is limited to
finite-temperature but can be made
very precise by increasing the number of time slices and the number of Monte-Carlo
sweeps (we used 128 slices in imaginary time and up to $5\times10^5$
Monte Carlo sweeps in practice). Also, using QMC allows for a reconstruction of
the spectral functions using a numerical analytic continuation based on
the maximum entropy algorithm.

\begin{figure}[h]
\includegraphics[width=8cm,]{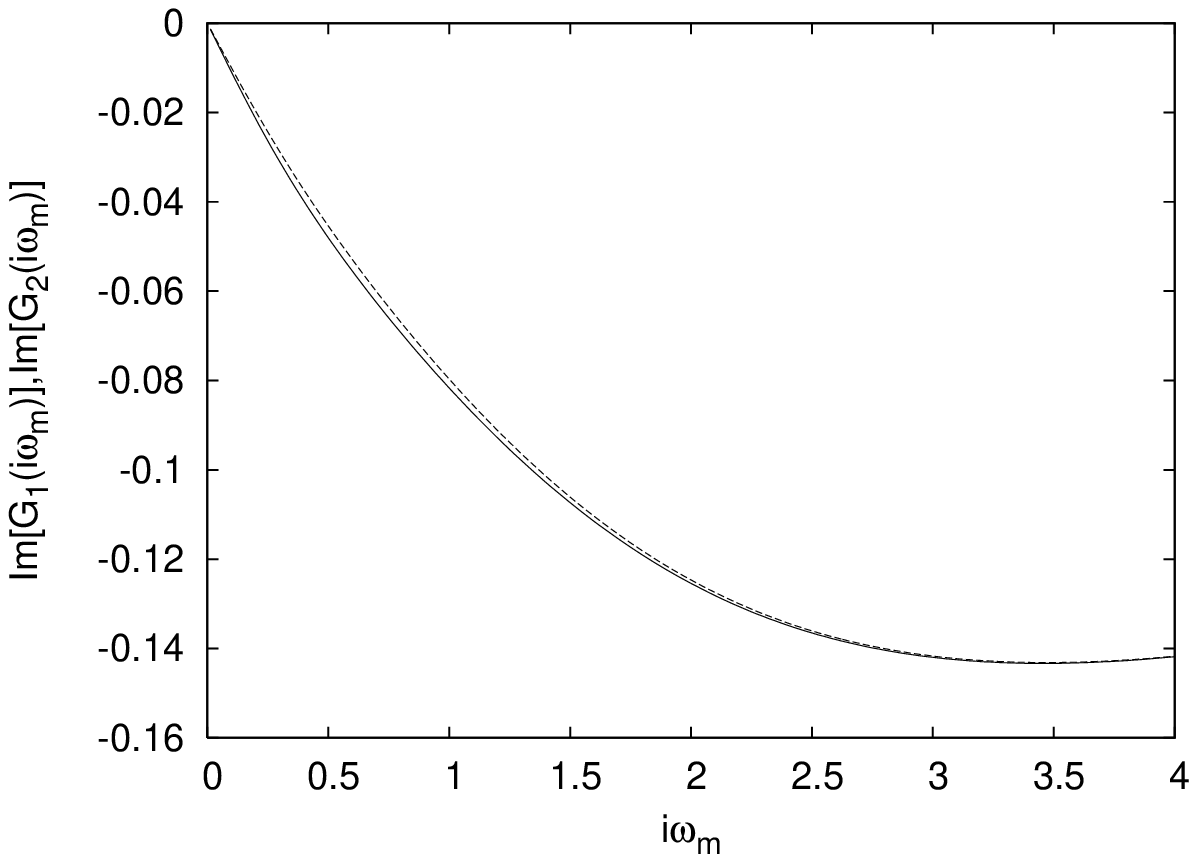}
\includegraphics[width=8cm,]{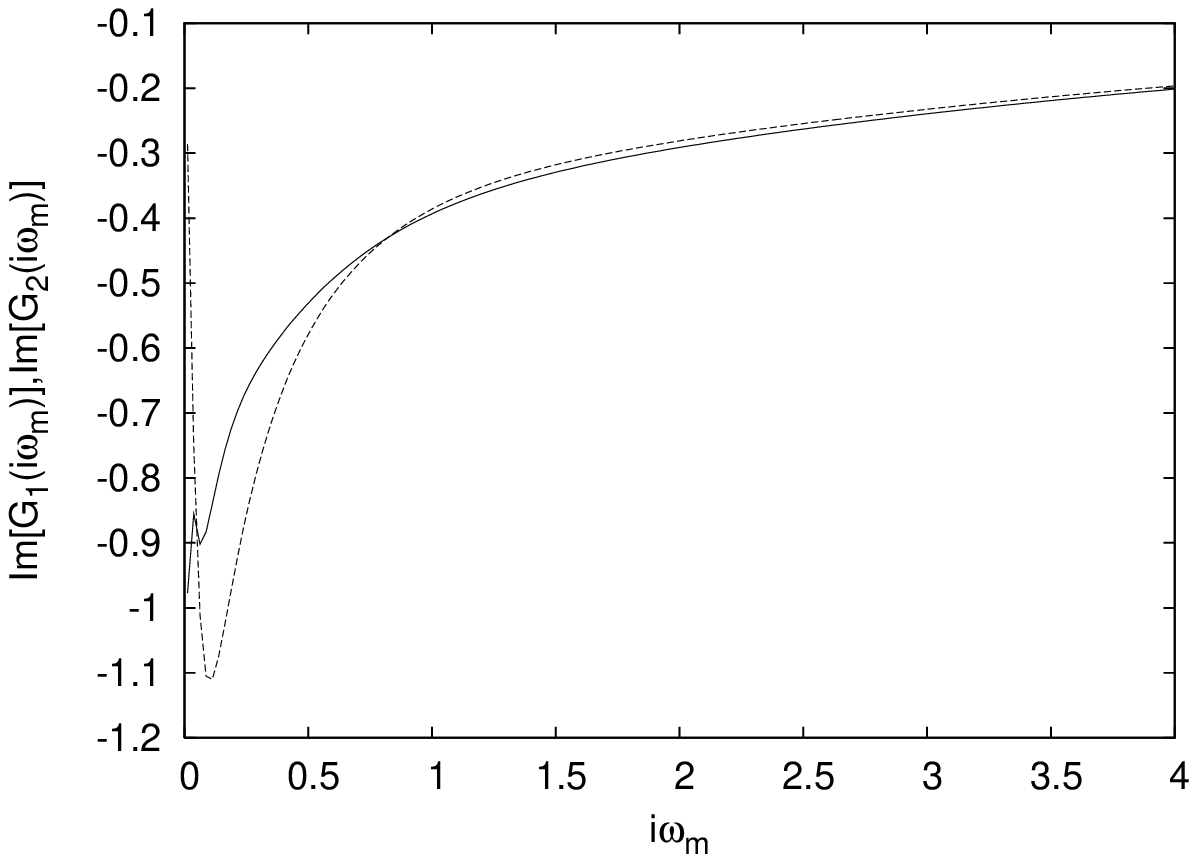}
\includegraphics[width=8cm,]{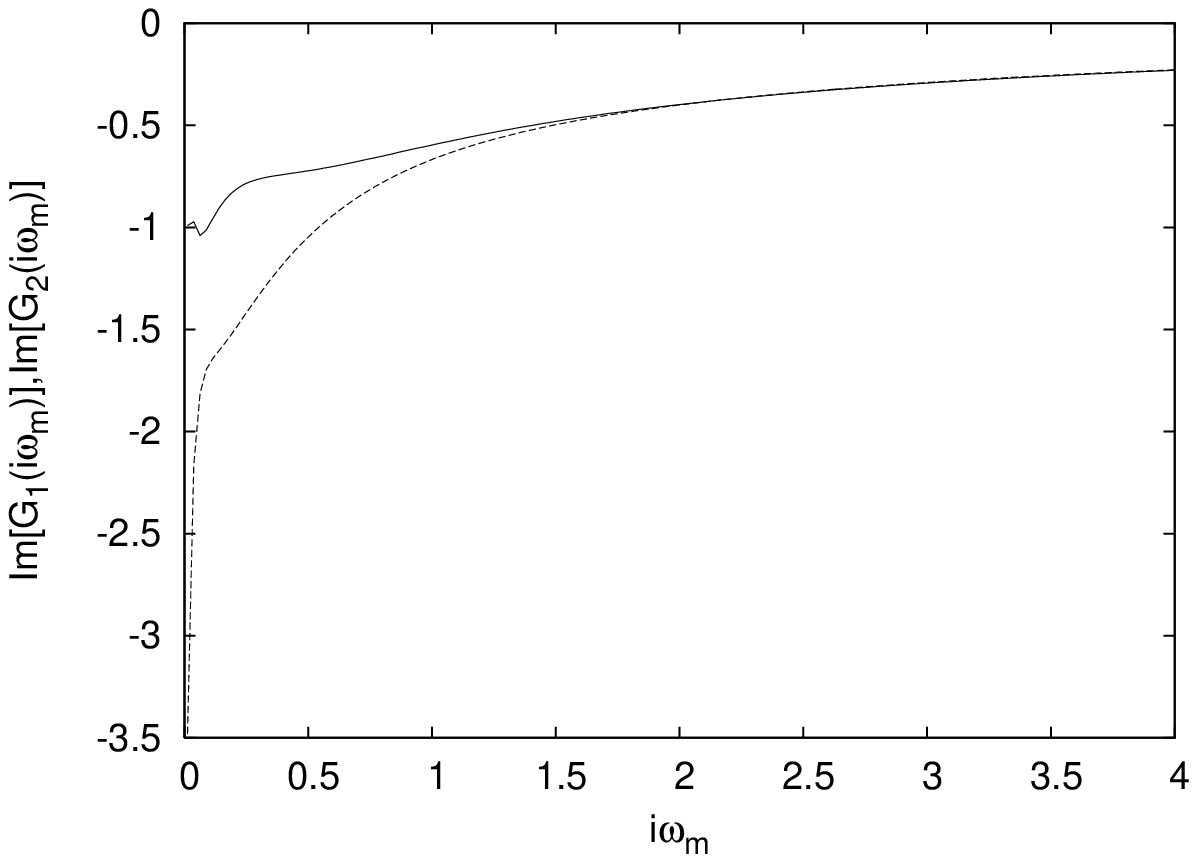}
\caption{\label{fig:Green} Imaginary part of the Green functions of the two bands at low energy
for $t_2/t_1=0.16$. Upper panel: $U/t_1=7.0$, both bands are insulating. Central panel: $U/t_1=4.0$,
Orbital Selective Mott phase. Lower panel: $U/t_1=2.0$, both bands metallic.
The dashed curves correspond to the orbital with narrower bandwidth.
All energy scales are in units of $t_1$.}
\end{figure}

In Fig.~\ref{fig:Green}, we display the ED results for the local Green's functions
on the Matsubara axis, for a small bandwidth ratio $t_2/t_1$ and three
different values of $U$ corresponding to the insulating, metallic, and orbital-selective
Mott phases. In the particle-hole symmetric case, the Green's functions are
purely imaginary on the Matsubara axis and related to the spectral function $A_m(\epsilon)$
of each orbital by:
\be
\label{eq:spectral}
{\rm Im} G_m(i\omega) = -2\omega\,\int_0^{+\infty}
d\epsilon\, \frac{A_m(\epsilon)}{\omega^2+\epsilon^2}
\ee

When the spectral function $A_m(\epsilon)$ has a gap, the integral in the
right-hand side of (\ref{eq:spectral}) has no singularity in the
$\omega\rightarrow 0$ limit, and hence ${\rm Im} G_m(i\omega)\propto\omega$ at low
frequency. Furthermore, ${\rm Im} G_m(i\omega)$ has a minimum for $\omega$ of
order $\Delta_m/2$, with $\Delta_m$ the gap in the m-th orbital (as can be seen
by replacing $A_m(\epsilon)$ by the simplified form
$1/2[\delta(\epsilon-\Delta_m/2)+\delta(\epsilon+\Delta_m/2)]$, yielding
${\rm Im} G_m(i\omega)\simeq -2\omega/(\omega^2+\Delta_m^2/4)$).
This is fully consistent with the ED results in the upper panel of
Fig.~\ref{fig:Green}, corresponding to a large value of $U/t_1=7$,
with both orbitals insulating and having a gap of order $U$.

In the lower panel of Fig.~\ref{fig:Green}, ED results are displayed for
$U/t_1=2$, when both orbitals are metallic. In this case, the low-frequency
limit of (\ref{eq:spectral}) yields: ${\rm Im} G_m(i\omega\rightarrow 0)=-\pi A_m(0)$.
This is consistent with the numerical results, which also show that the Luttinger
theorem is obeyed for both bands: $\pi A_m(0)=1/t_m$.

The central panel of Fig.~\ref{fig:Green} displays the ED results for
an intermediate coupling, corresponding to the orbital-selective phase.
One sees that the wider band has metallic behaviour, with $A_1(\omega)$
still reaching the Luttinger value at $\omega=0$. In contrast,
${\rm Im} G_2(i\omega)$ appears to vanish as $\omega\rightarrow 0$, within the
energy resolution of ED. However, in striking contrast to the upper panel (insulating
phase), the minimum in ${\rm Im} G_2(i\omega)$ is at a {\it very low frequency scale}
which is obviously not given by $U$. This strongly suggests that $A_2(\omega)$ displays
low-energy peaks very close to $\omega=0$, and may even have spectral weight down to
arbitrary low frequency.

\begin{figure}
\includegraphics[width=8cm]{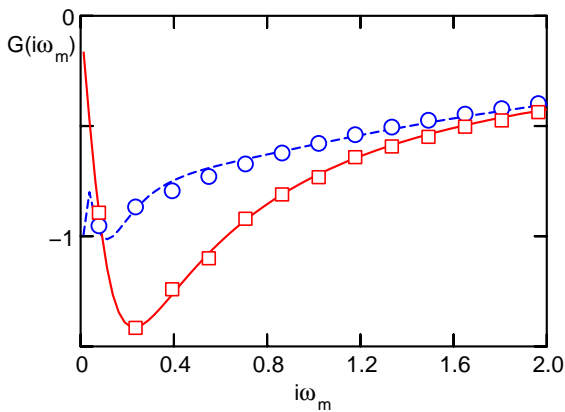}
\caption{\label{fig:Green_compared} (color online). Imaginary
parts of the Green's functions in Matsubara space for $t_2/t_1
=0.1$, $U/t_1=1.6$. Solid and dashed lines represent the exact
diagonalization results for the wide and narrow bands
respectively. Circles and squares represent QMC data for the same
quantities at $\beta t_1=40$.}
\end{figure}

\begin{figure}
\includegraphics[width=8cm]{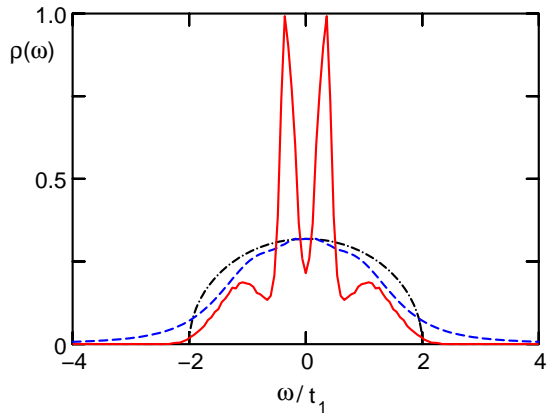}
\caption{\label{fig:dos} (color online). Spectral function for
$t_2/t_1 =0.1$, $U/t_1=1.6$ at $\beta t_1=40$ in QMC-DMFT. The
solid and dashed red lines denote the narrow and wide bands
respectively. The DOS of the non-interacting system (wide band
only) is given for comparison.}
\end{figure}

Figure \ref{fig:Green_compared} compares the ED and QMC
results for ${\rm Im} G_{1,2}(i\omega)$ in the
strongly anisotropic case $t_2/t_1 =0.1$ for a relatively
small Coulomb interaction $U/t_1=1.6$, easier to study
with QMC. These parameters
also correspond to the orbital-selective phase (Fig.~\ref{fig:DMFT}).
Very good agreement between the two methods is found, confirming the
above analysis (and confirming also the Luttinger value for the
wider band with greater accuracy than in ED). The corresponding
spectral functions obtained by the maximum-entropy method are
displayed on Fig.~\ref{fig:dos}. The spectral function of
the broader band is only slightly modified as compared to the
non-interacting d.o.s. Small shoulders are visible, at the position of
the lower and upper Hubbard bands, the Luttinger theorem is obeyed, and
some of the spectral weight is transfered to higher energies as expected.
The narrow-band however is obviously in a strong-coupling regime, with well-marked
upper and lower Hubbard bands. The most striking features however are the
two narrow peaks at low-frequency, which can be interpreted either
as a split quasi-particle resonance or as the sign of a {\it pseudo-gap}
(partially filled by thermal excitations since the QMC calculation is
for $T/t_1=1/40$).

Hence, the general conclusion of this analysis is that the orbital-selective
phase found for $J=0$ at small enough $t_2/t_1$ is not a conventional Mott phase
in which the (``localized'') orbital with narrower bandwidth would display a sharp gap.
Instead, two narrow peaks exist near $\omega=0$ and
finite spectral weight is found down to low-energy. Our numerical data
are consistent with a pseudo-gap behaviour, but a precise characterization of the
low-energy nature of the phase will require further effort, using highly precise
techniques at low-energy such as the numerical renormalization group. This is left for
future work. Such a study should also clarify in which precise sense the narrower band is
``localized'' in this phase, and whether the orbital-selective transition is a true
phase transition or rather a sharp crossover.

\section{Instability of the orbital-selective Mott phase upon
inter-band hybridization}
\label{sec:hyb}

To what extent an orbital-selective Mott phase may occur in
practice depends on its stability with respect to perturbations.
This is an important issue in view of the fact that the
hamiltonian considered in this paper has a rather high degree of
symmetry. In this last section, we consider the effect of an
hybridization between the two orbitals (also recently considered
in Ref.~\cite{koga_0503651}), i.e of a local non-diagonal term:
\be \label{eq:hyb} H_{hyb}=V\sum_{i\s}
(d^\dagger_{i1\s}d_{i2\s}+d^\dagger_{i2\s}d_{i1\s}) \ee We note
that this term could be eliminated by diagonalizing the
non-interacting hamiltonian. However, in the new basis, the
interaction terms will be modified: terms will be generated which
will have the same physical effect than a hybridisation (and will
involve non-local contributions in general). Indeed, as emphasized
in Ref.~\cite{demedici_mottpam_2005}, the existence of an OSMT is
a {\it basis- independent} issue. In a general two-band model, a
Mott transition is signaled, when approached from the metallic
side, by a low-frequency singularity in
$\omega\hat{I}-\hat{\Sigma}(\omega)=\hat{Z}^{-1}\omega +\cdots$,
where  $\hat{\Sigma}$ and $\hat{Z}$ are the self-energy and
quasiparticle-weight matrices, respectively. An OSMT is
characterized by $\hat{Z}$ having one zero-eigenvalue while the
other one remains finite. Being associated with the rank of the
$\hat{Z}$-matrix, it is a basis-independent notion. Our choice of
basis is such that the interaction terms have the form specified
above.

\subsection{Physical considerations}

Some statements about the effect of a finite hybridization can be made on general physical
grounds (focusing for simplicity on the $J=0$ case).
First, for small values of $U$ when both orbitals are itinerant, it is
obvious that a hybridization will not change qualitatively the low-energy nature of the metallic
phase. Also, at very large $U$, when both orbitals are localized and a gap exists in both
orbital sectors, we expect the presence of a gap to be a robust feature which
persists in the presence of $V$. However, it is also clear that introducing $V$ into
this gapped insulator allows the local moments formed in the Mott insulating
state at $V=0$ to screen each other. This occurs through the formation of on-site
bonding and antibonding ``molecular'' levels mixing the two orbitals.
As a result, the local-moment Mott insulator is
expected to be replaced by a Mott insulator in which {\it local singlets} are formed on
each site. The intermediate $U$ regime, in which the system is in the OSMP at $V=0$, is more
delicate. Because orbital 1 is itinerant, a Kondo screening process can take place, which will
screen the local moment (formed by orbital 2 when $V=0$).
The resulting state can a priori be either a heavy-fermion metallic state involving
quasiparticles with a large effective mass, or
(because we are considering the half-filled case), a Kondo insulating state in which a gap is
formed in the low-energy quasiparticle spectrum.
Below, we study this question using the slave-spin
approach and find that both phases can be obtained, depending on the value of $U$ and $V$.
A larger $V$ favours the opening of a gap, as expected.
Using a general low-frequency analysis, we also demonstrate that, for small values of $V$,
the heavy-fermion metallic state, and not the Kondo insulating phase, is induced.
We note, finally, that both the Kondo-insulating phase, and the phase obtained at large $U$
in which on-site local moments in different orbitals screen each other,
have a singlet ground-state. It is
therefore not obvious a priori whether these two phases are continuously
connected~\cite{koga_0503651}, as
found in \cite{schork_prb_1997,sato_jpsj_2004} for a related (but different) model, or whether
a phase transition between them can exist. We return to this point at the end of
this section.

In any event, these physical arguments imply that the orbital-selective Mott phase is
unstable with respect to the introduction of a non-zero hybridization, at zero-temperature.
Of course, for temperatures above the Kondo scale, the physics of the OSMP can be recovered. This
is an important point in view of the possible experimental relevance of the OSMP.

\subsection{Slave-spin mean-field study}

We now turn to a more quantitative study of the effect of a finite hybridization
(\ref{eq:hyb}), using the slave-spin mean-field approximation.
Our aim is not to establish a full phase diagram for all values of the parameters
$U$, $J$ and $V$, but rather to investigate whether this approach does support
the physical expectations discussed above. A more extensive
investigation will be presented in a future publication.

\begin{figure}[h]
\includegraphics[width=8cm]{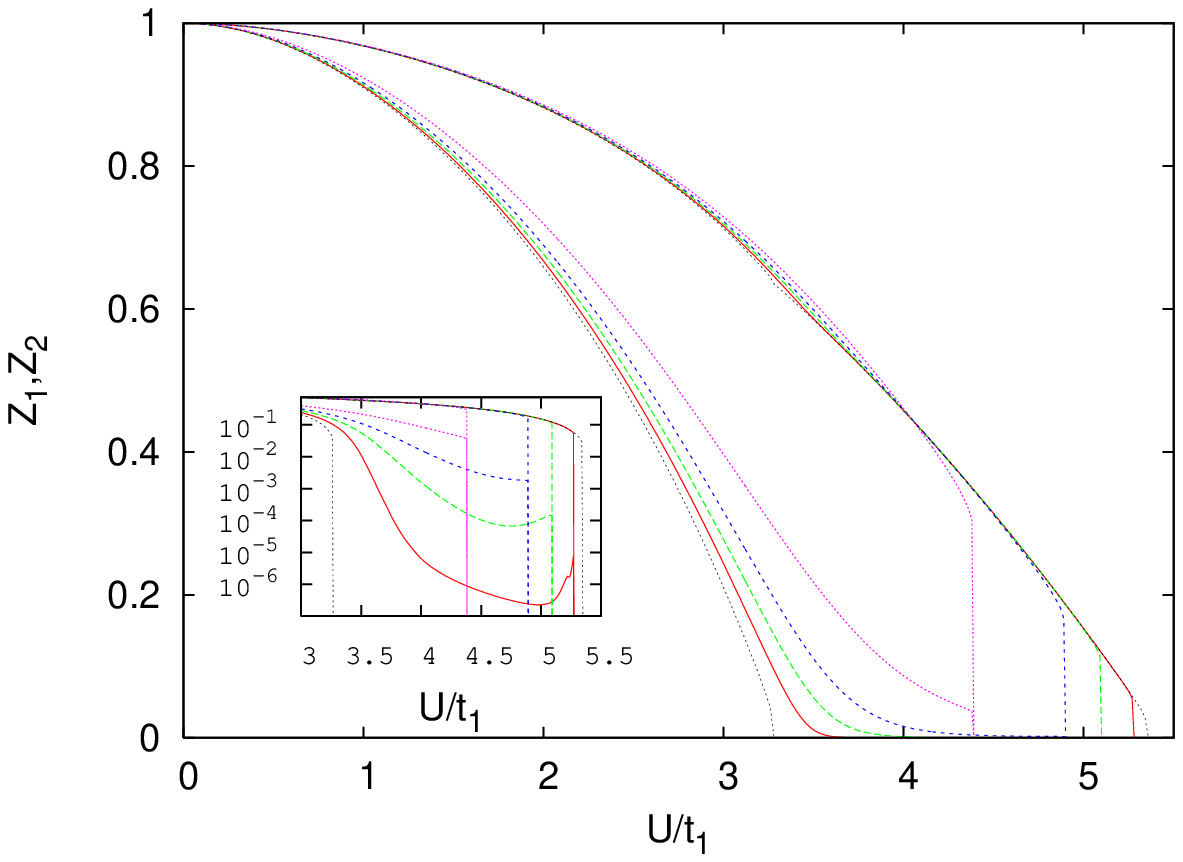}
\includegraphics[width=8cm]{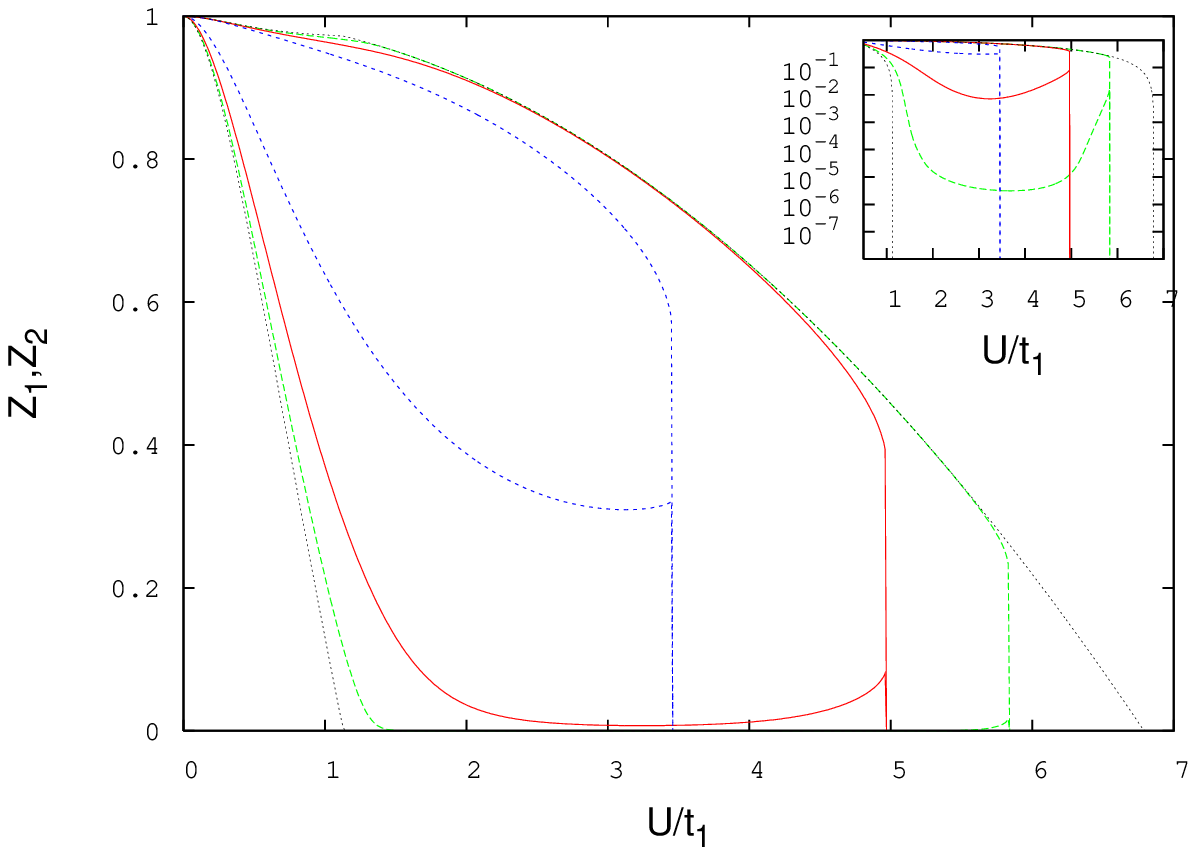}
\caption{\label{fig:hybridization} (color online). Quasiparticle
residues $Z_1$ and $Z_2$ within slave spins MFT, for finite $V$.
Top: $t_2/t_1=0.5$,$J=0.25U$ for (from right to left in the upper
manifold -wide band -, from left to right in the lower manifold -
narrow band): $V=0$,(OSMT system),$V/t_1=0.1,0.15,0.2,0.3$.
Bottom: $t_2/t_1=0.1$,$J=0$ for (from right to left in the upper
manifold -wide band -, from left to right in the lower manifold -
narrow band) $V=0$,(OSMT system),$V/t_1=0.05,0.1,0.2$. Insets show
the same graph in log scale.}
\end{figure}

In Fig.~\ref{fig:hybridization}, we display the quasiparticle residues $Z_1$ (for
the broad band) and $Z_2$ (for the narrow band), as a function of $U$ and for increasing
values of the hybridization $V$. It is seen that, starting from the OSMP in which
$Z_1\neq 0$ and $Z_2=0$ for $V=0$, one obtains either a phase in which both
$Z_1$ and $Z_2$ are non-zero (i.e Kondo screening takes place), or an insulating phase in
which $Z_1=Z_2=0$. This demonstrates that the
hybridization is indeed a singular perturbation on both the $J=0$ and finite-$J$
orbital-selective Mott phases, in agreement with the physical arguments above.

\begin{figure}[h]
\includegraphics[width=8cm]{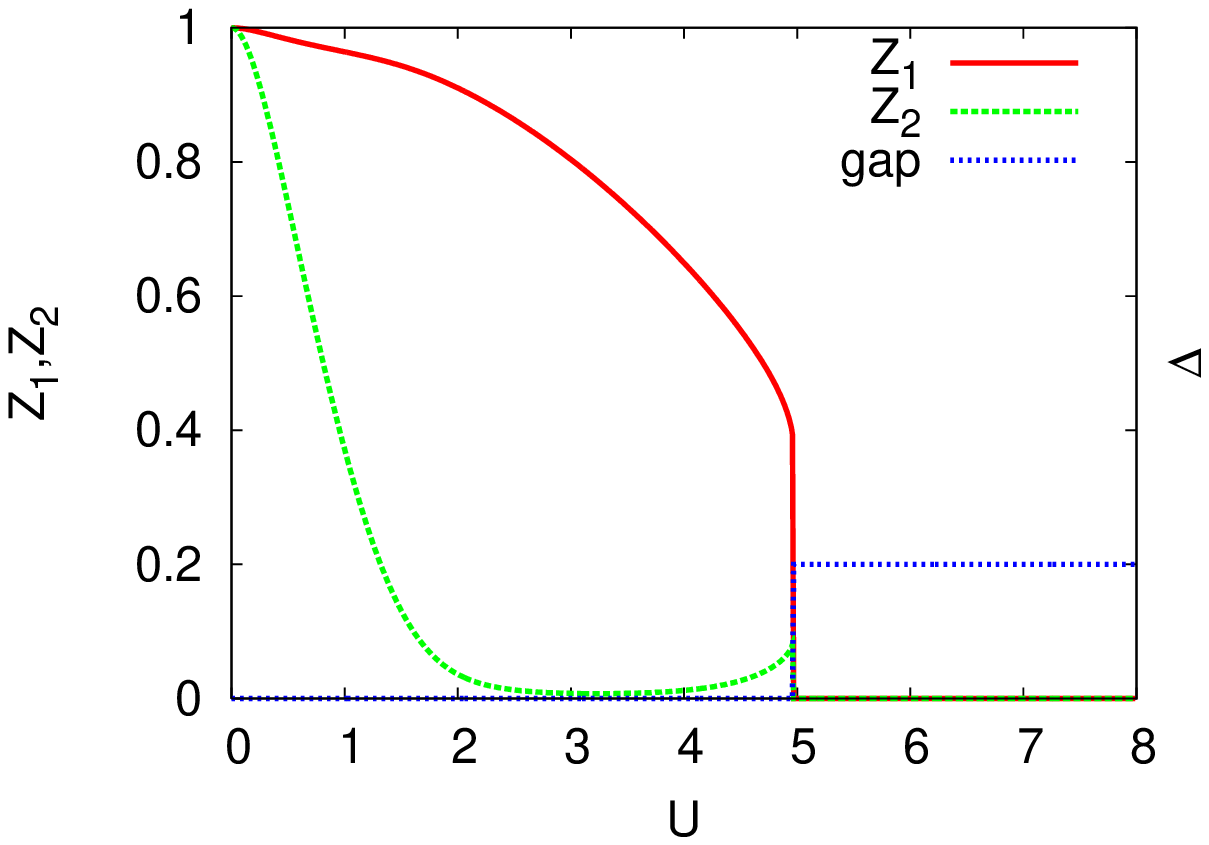}
\includegraphics[width=8cm]{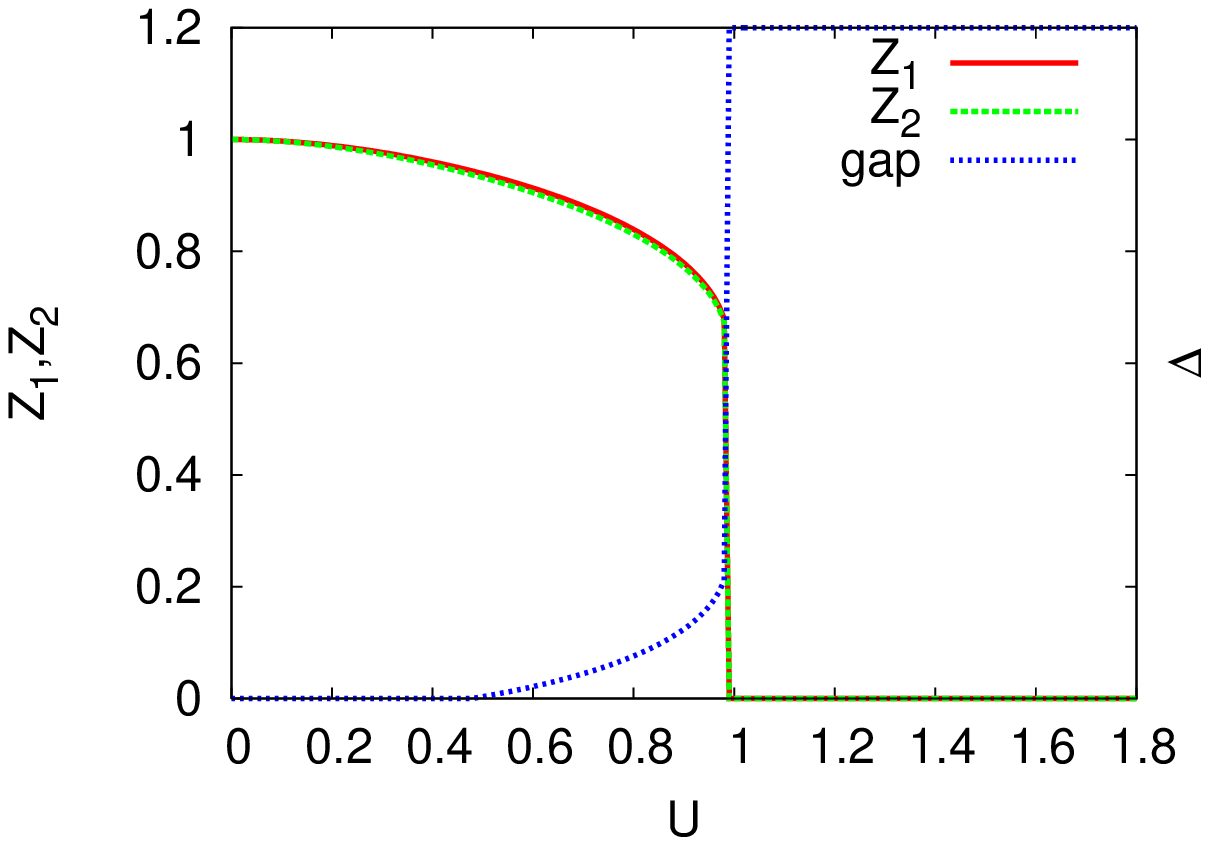}
\caption{\label{fig:bandgap} (color online). Gap amplitude of the
auxiliary fermions (see text) and quasiparticle residues within
slave spins MFT, for two values of $V$. Top: $t_2/t_1=0.1$,$J=0$
for $V/t_1=0.1$ (upper panel), and $V/t_1=0.6$ (lower panel). The
corresponding critical V for the non-interacting system is
$\sqrt{D_1D_2}\simeq 0.63$ in this case.}
\end{figure}

In order to check whether the Kondo-screened phase is a
(heavy-fermion) metal or whether it is gapped (Kondo insulator),
we have plotted in Fig~.\ref{fig:bandgap} the band gap of the
auxiliary quasiparticles found within slave spin mean-field theory
(the plot is for $J=0$, the $J\neq 0$ case being similar). Note
that due to the factorization of the Green's function
\begin{equation}
\langle  d(\tau)_{m\sigma} d^{\dagger}_{m\sigma}(0) \rangle = 4
\langle S^x(0)_{m\sigma} S^x(\tau)_{m\sigma} \rangle \langle
f_{m\sigma}(\tau)f_{m\sigma}^{\dagger}(0) \rangle
\end{equation}
the physical gap in the insulating phase will be of the order $U$.
For small and intermediate values of $V$, the phase with $Z_1\neq
0$ and $Z_2\neq 0$ is metallic (gapless) (see
Fig~.\ref{fig:bandgap}). The orbital-selective Mott phase is
replaced by a heavy- fermion regime, as shown in Fig.
\ref{fig:hybridization}, in which the orbital with narrower
bandwidth acquires a very large effective mass (corresponding to a
very low quasiparticle coherence scale $Z_2$). This is also in
qualitative agreement with our recent study of the periodic
Anderson model with direct f-electron
hopping~\cite{demedici_mottpam_2005}. Only beyond a critical value
of $V$ is a gapped Kondo insulator found. As discussed below, this
can in fact be proven generally, beyond the mean-field
approximation used here.

At small to intermediate values of $V$, the two Mott transitions associated with the
OSMP at $V=0$ are therefore replaced by a single non-selective
transition from a (heavy-fermion) metal to an insulator
as $U$ is increased. Within the slave-spin mean-field, this unique metal to insulator transition is
found to be first-order, and to occur at a critical value of $U$ which lies in between the
critical interactions
of the metal-OSMP and OSMP-insulator transitions.

For larger values of $V$, the OSMP phase is replaced by a Kondo-insulating phase
with $Z_1,Z_2\neq 0$ but a finite quasiparticle band- gap (lower plot in
Fig.~\ref{fig:hybridization}). Within slave-spin mean-field,
a phase transition takes place as $U$ is increased, towards another insulating phase
with $Z_1=Z_2=0$ (corresponding to the fact that the Kondo effect does not take place when
$V$ is turned on starting from a Mott phase for both orbitals, with a large gap).
As pointed out above however, this large-$U$ insulating phase also has a singlet
ground-state however, due to inter-orbital screening.
Whether this phase transition is an artefact of the slave-spin mean-field or whether
it is indeed present in a more accurate DMFT treatment is a question to which we shall return
below.

\subsection{General low-frequency analysis}

In order to understand better the nature of both the metallic and the
insulating phases to which the OSMP is driven for $V\neq 0$, we perform here
a low-energy analysis in terms of a renormalized quasiparticle
band-structure (see also \cite{sato_jpsj_2004}).
We focus on the case in which the Kondo effect does take place as $V$ is
turned on, resulting in both quasiparticle residues being finite.
We thus keep the following
terms in the low-frequency expansion of the self-energy:
\begin{eqnarray}
&\Sigma_{11}(\omega) = \omega (1-1/Z_1)+\cdots\nonumber \\
&\Sigma_{22}(\omega) = \omega (1-1/Z_2)+\cdots\nonumber \\
&\Sigma_{12}(\omega) = \Sigma_{12}(0)+\cdots
\end{eqnarray}
The low-energy quasiparticle band structure is then given by:
\begin{equation}
[\omega-Z_1\epsilon_1(\vk)][\omega-Z_2\epsilon_2(\vk)]-Z_1Z_2 [V+\Sigma_{12}(0)]^2=0
\end{equation}
From this expression, it is easily seen that a gap is present whenever:
\begin{equation}
\veff\,>\,\sqrt{\deff1 \d2eff}
\label{eq:cond_gap}
\end{equation}
In this expression, $\veff$ is the effective hybridisation between low-energy
quasiparticles, and $\deff1,\d2eff$ the renormalised quasiparticle half-bandwidths,
given by:
\begin{equation}
\veff=\sqrt{Z_1Z_2}[V+\Sigma_{12}(0)]\,\,,\,\,
\deff1=Z_1 D_1\,\,,\,\,\d2eff=Z_2D_2
\label{eq:eff}
\end{equation}
Remarkably, $Z_1$ and $Z_2$ drop out from (\ref{eq:cond_gap}), and the
condition for a quasiparticle bandgap therefore reads:
\begin{equation}
V+\Sigma_{12}(0)\,>\,\sqrt{D_1D_2}
\label{eq:cond_gap_dropZ}
\end{equation}
If this condition is satisfied (and $Z_1,Z_2\neq 0$), one has a Kondo insulator phase,
with a quasiparticle band gap given by:
\begin{equation}
\Delta_g = \sqrt{(\deff1-\d2eff)^2+4\veff^2}- (\deff1+\d2eff)
\end{equation}
In the opposite case, a heavy-fermion (gapless) metallic phase is formed.

From (\ref{eq:cond_gap_dropZ}), it is seen that it is the off-diagonal component of
the self-energy (induced by $V$) which plays the key role in deciding whether
a gap opens or not. From the same equation, it is also clear that {\it a band-gap cannot open
for arbitrary small $V$}. Indeed, at small $V$, $\Sigma_{12}$ grows at most proportionally
to $V$, and hence the left-hand side of (\ref{eq:cond_gap_dropZ}) is small, while the
right-hand side is finite (note that the r.h.s involves the {\it bare} bandwidths).
Hence, a critical value of $V$ is required to open a band gap and
enter the Kondo insulating phase, starting from the OSMP. Note that this analysis is
general and relies only on Fermi-liquid considerations, independently of the specific
method used to solve the model. Note also that the situation is very different when one
of the bare bands is dispersionless ($D_2=0$), as is the case of the periodic Anderson models
considered in \cite{schork_prb_1997,sato_jpsj_2004}. In this case, it is clear from
(\ref{eq:cond_gap_dropZ}) that an arbitrarily small $V$ induces the Kondo insulating state.
The formation of a heavy-fermion metallic state induced by a non-zero hybridization
in a generalized periodic Anderson model with $D_2\neq 0$ (direct f-electron hopping)
was also investigated in our recent work~\cite{demedici_mottpam_2005}.

The slave-spin mean field results presented above can be placed in the context of this
general low-frequency analysis. Within this approach, one has:
\begin{equation}
Z_1=4\langle S_1^x\rangle^2\,\,,\,\,
Z_2=4\langle S_2^x\rangle^2\,\,,\,\,
\veff = 4 \langle S_1^x S_2^x\rangle\,V
\end{equation}
It is important to realize that this approach does provide an off-diagonal
component of the physical electron self-energy, which, using
(\ref{eq:eff}), is given by:
\begin{equation}
\frac{\Sigma_{12}(0)}{V}\,=\,
\frac{\langle S_1^x S_2^x\rangle - \langle S_1^x\rangle\langle S_2^x\rangle}
{\langle S_1^x\rangle\langle S_2^x\rangle}
\end{equation}
Note that in more conventional slave-boson approaches, the physical
electron operators are related to the quasiparticles by a slave-boson condensation
amplitude which is a c-number ($d^\dagger_m = \sqrt{Z_m}f^\dagger_m$). As a result,
$\veff=\sqrt{Z_1Z_2} V$, and no off-diagonal component of the self-energy is present.
In such slave-boson approaches, the condition for the presence of a quasiparticle
band gap is therefore entirely unrenormalized by interactions and reads:
$V/\sqrt{D_1D_2}>1$. This is an oversimplification which is not present in the
slave-spin approach. There, the criterion for the opening of a gap reads:
\begin{equation}
\frac{V}{\sqrt{D_1D_2}}\,>\,
\frac{\langle S_1^x\rangle\langle S_2^x\rangle}{\langle S_1^x S_2^x\rangle}
\end{equation}
However, despite the renormalisation of the hybridisation by the off-diagonal
self-energy, we find that in practice this criterion is very close to the
non-interacting one, which gives a reasonable approximation of the
critical hybridization necessary to open a band gap.
We also note that, within the slave-spin approximation, the insulating gap
takes its bare value $2V$ (corresponding to the splitting between the bonding
and antibonding molecular levels) as soon as $Z_1=Z_2=0$, which is certainly an
over-simplification.

Finally, we comment on the phase transitions between the different phases
induced by a non-zero hybridization. At small $V$, it is clear that there must
exist a phase transition between the metallic (gapless) heavy-fermion phase
and the insulating (gapped) phase, as $U$ is increased (as indeed seen on
Fig.~\ref{fig:bandgap}, top panel).
The situation is less clear at larger $V$. There, we always expect an insulator with a
singlet ground-state. However, the mechanism behind this singlet formation is rather
different at smaller and larger values of $U$. In the former case, Kondo screening dominates
and the singlet is formed by screening the local moment in orbital 2 by the electrons in
orbital 1. At larger $U$, the Kondo coupling is smaller, and it is the formation of
an inter-orbital molecular bonding level (due to $V$) which is responsible for the
singlet formation. Within slave-spin mean theory, we find a phase transition between these
two kinds of insulator as $U$ is increased, the latter one being signalled by
$Z_1=Z_2=0$ (Fig.~\ref{fig:bandgap}, bottom panel). However, it is not clear whether
this phase transition is a real feature or an artefact of the slave-spin approximation.
In \cite{koga_0503651}, Koga et al. recently suggested that these two phases should be
adiabatically connected, as found also in the study of the periodic Anderson model
with correlated conduction electrons~\cite{schork_prb_1997,sato_jpsj_2004} (note however
that the present model is different, in that both bands have a dispersion and that an
inter-orbital interaction is present). The problem is qualitatively similar (but not equivalent, because
of the inter-orbital interaction) to the
two-impurity Kondo problem~\cite{2impKondo}, in which the inter-impurity singlet (RKKY) fixed points and the
Kondo- singlet fixed points are in general adiabatically connected (except in the special case
of particle-hole symmetry, where a phase transition does occur). Note that it is well known~\cite{2impslave}
that for this latter model, slave boson approximations do lead to spurious first-order
transitions.
A full answer to this question is beyond the scope of this paper, and is left for
a future investigation, together with a complete phase diagram as a function of interaction
strength and hybridization.

\section{Conclusion}

In this article, we have studied whether the Mott transition of a half-filled,
two-orbital Hubbard model with unequal bandwidths occurs simultaneously for both bands
or whether it is a two-stage process in which the orbital with narrower bandwidth localizes first (giving
rise to an intermediate `orbital-selective' Mott phase).
In order to study this question, we have used two techniques. The first is a
mean-field theory based on a new representation of fermion operators in terms of
slave quantum spins. This method is similar in
spirit to the Gutzwiller approximation, and the slave-spin representation has a rather
wide range of applicability to multi-orbital models. The second method is dynamical mean-field theory,
using exact diagonalization and Quantum Monte-Carlo solvers.

The results of the slave-spin mean-field confirms several aspects of previous
studies~\cite{Koga_OSMT,Koga_OSMT_SCES}, and in particular the possibility of
an orbital-selective Mott transition.
However, some of the
conclusions differ from those of previous work. Specifically, the slave-spin
approximation suggests that
a critical value of the bandwidth ratio $(t_2/t_1)_c$ exists, such that
the Mott transition is orbital-selective for arbitrary value of the Coulomb exchange
(Hund coupling) $J$ when $t_2/t_1<(t_2/t_1)_c$. When $t_2/t_1>(t_2/t_1)_c$, $J$ has to
be larger than a finite threshold for an OSMT to take place. This suggests that
the existence of an OSMT is not simply related to the symmetry of the interaction term only.
In particular, an intermediate phase is found for $J=0$ at small $t_2/t_1$.

We have studied whether DMFT confirms these findings, and found that the main qualitative
conclusions on the existence of the orbital-selective phase are indeed the same, but that
the nature of the intermediate phase at $J=0$ is a rather subtle issue. Indeed, the narrow band
does not have the properties of a gapped Mott insulator in this phase and displays finite spectral
weight down to arbitrary low-energy. This is, for example, consistent with a pseudo-gap
behaviour but requires further studies to be fully settled (using e.g low-energy techniques
such as the numerical renormalization group).

We note also that our study emphasizes the key role of the exchange and (on-site) inter-orbital
pair hopping terms in the Coulomb hamiltonian in stabilizing the orbital-selective phase, in
agreement with Koga et al.~\cite{Koga_OSMT_SCES}.

Finally, we found that the orbital-selective Mott phase is
generically unstable with respect to an inter-orbital
hybridization $V$. In the presence of such a term, two possible
phases are obtained, depending on the strength of $U$ and $V$.
Either the narrow orbital acquires a large (but finite) effective
mass, corresponding to a heavy-fermion metallic state. Or the
system is an insulator with a gap. This insulator differs from the
Mott insulator at $V=0$ since it has a singlet ground-state. This
is due to screening processes, involving both Kondo exchange and
the formation of an on-site molecular (bonding) level. Whether one
has in fact two different insulating phases separated by a phase
transition (each phase being dominated by one of these screening
processes) -as obtained by slave-spin mean-field-, or whether one
has a simple crossover~\cite{koga_0503651} is an open question
which deserves further study.

Of course, at intermediate temperature (above the quasiparticle
coherence scale of the narrower band, but below that of the wider
band), a physics similar to the orbital-selective Mott phase can
be recovered even in the presence of a finite hybridization. This
orbital-selective heavy-fermion state might be relevant to the
physics of Ca$_{2-x}$Sr$_x$RuO$_4$. This is indeed supported by
the recent angular magnetoresistance oscillations experiments of
Balicas et al.~\cite{balicas_amro_2005}.

\smallskip
\acknowledgments
During the completion of this paper, we learned of the work by M.~Ferrero, F.~Becca, M.~Fabrizio and
M.~Capone, reaching similar conclusions. We are grateful to F.~Becca, M.~Fabrizio and
M.~Capone for discussions.
A.G acknowledges discussions with S.~Florens and N.~Dupuis on the slave-spin representation, at
an early stage of this work. We are grateful to A.~Lichtenstein for help with ED calculations in the
multi-orbital context. We also thank
A.~Koga, N.~Kawakami, G.~Kotliar, T.M.~Rice and M.~Sigrist for useful discussions.
Finally, we thank the referees for their constructive comments.
This research was supported by CNRS and Ecole Polytechnique
and by a grant of supercomputing time at IDRIS Orsay (project 051393).

\bibliographystyle{apsrev}

\begin{thebibliography}{23}
\expandafter\ifx\csname natexlab\endcsname\relax\def\natexlab#1{#1}\fi
\expandafter\ifx\csname bibnamefont\endcsname\relax
  \def\bibnamefont#1{#1}\fi
\expandafter\ifx\csname bibfnamefont\endcsname\relax
  \def\bibfnamefont#1{#1}\fi
\expandafter\ifx\csname citenamefont\endcsname\relax
  \def\citenamefont#1{#1}\fi
\expandafter\ifx\csname url\endcsname\relax
  \def\url#1{\texttt{#1}}\fi
\expandafter\ifx\csname urlprefix\endcsname\relax\def\urlprefix{URL }\fi
\providecommand{\bibinfo}[2]{#2}
\providecommand{\eprint}[2][]{\url{#2}}

\bibitem[{\citenamefont{{Georges} et~al.}(1996)\citenamefont{{Georges},
  {Kotliar}, {Krauth}, and {Rozenberg}}}]{georges_review_dmft}
\bibinfo{author}{\bibfnamefont{A.}~\bibnamefont{{Georges}}},
  \bibinfo{author}{\bibfnamefont{G.}~\bibnamefont{{Kotliar}}},
  \bibinfo{author}{\bibfnamefont{W.}~\bibnamefont{{Krauth}}}, \bibnamefont{and}
  \bibinfo{author}{\bibfnamefont{M.~J.} \bibnamefont{{Rozenberg}}},
  \bibinfo{journal}{Reviews of Modern Physics} \textbf{\bibinfo{volume}{68}},
  \bibinfo{pages}{13} (\bibinfo{year}{1996}).

\bibitem[{\citenamefont{{Georges}}(2004)}]{georges_strong}
\bibinfo{author}{\bibfnamefont{A.}~\bibnamefont{{Georges}}}, in
  \emph{\bibinfo{booktitle}{Lectures on the physics of highly correlated
  electron systems VIII}}, edited by
  \bibinfo{editor}{\bibfnamefont{A.}~\bibnamefont{Avella}} \bibnamefont{and}
  \bibinfo{editor}{\bibfnamefont{F.}~\bibnamefont{Mancini}}
  (\bibinfo{publisher}{American Institute of Physics}, \bibinfo{year}{2004}),
  \bibinfo{note}{cond-mat/0403123}.

\bibitem[{\citenamefont{Kotliar and
  Vollhardt}(2004)}]{kotliar_dmft_physicstoday}
\bibinfo{author}{\bibfnamefont{G.}~\bibnamefont{Kotliar}} \bibnamefont{and}
  \bibinfo{author}{\bibfnamefont{D.}~\bibnamefont{Vollhardt}},
  \bibinfo{journal}{Physics Today} \textbf{\bibinfo{volume}{March 2004}},
  \bibinfo{pages}{53} (\bibinfo{year}{2004}).

\bibitem[{\citenamefont{{Georges} et~al.}(2004)\citenamefont{{Georges},
  {Florens}, and {Costi}}}]{georges_mott_iscom}
\bibinfo{author}{\bibfnamefont{A.}~\bibnamefont{{Georges}}},
  \bibinfo{author}{\bibfnamefont{S.}~\bibnamefont{{Florens}}},
  \bibnamefont{and} \bibinfo{author}{\bibfnamefont{T.~A.}
  \bibnamefont{{Costi}}}, \bibinfo{journal}{Journal de Physique IV -
  Proceedings} \textbf{\bibinfo{volume}{114}}, \bibinfo{pages}{165}
  (\bibinfo{year}{2004}), \eprint{cond-mat/0311520}.

\bibitem[{\citenamefont{Anisimov et~al.}(2002)\citenamefont{Anisimov, Nekrasov,
  Kondakov, Rice, and Sigrist}}]{Anisimov_OSMT}
\bibinfo{author}{\bibfnamefont{V.}~\bibnamefont{Anisimov}},
  \bibinfo{author}{\bibfnamefont{I.}~\bibnamefont{Nekrasov}},
  \bibinfo{author}{\bibfnamefont{D.}~\bibnamefont{Kondakov}},
  \bibinfo{author}{\bibfnamefont{T.}~\bibnamefont{Rice}}, \bibnamefont{and}
  \bibinfo{author}{\bibfnamefont{M.}~\bibnamefont{Sigrist}},
  \bibinfo{journal}{Eur. Phys. J. B} \textbf{\bibinfo{volume}{25}},
  \bibinfo{pages}{191} (\bibinfo{year}{2002}).

\bibitem[{\citenamefont{Fang et~al.}(2004)\citenamefont{Fang, Nagaosa, and
  Terakura}}]{terakura_ruthenates}
\bibinfo{author}{\bibfnamefont{Z.}~\bibnamefont{Fang}},
  \bibinfo{author}{\bibfnamefont{N.}~\bibnamefont{Nagaosa}}, \bibnamefont{and}
  \bibinfo{author}{\bibfnamefont{K.}~\bibnamefont{Terakura}},
  \bibinfo{journal}{Phys. Rev. B} \textbf{\bibinfo{volume}{69}},
  \bibinfo{pages}{045116} (\bibinfo{year}{2004}).

\bibitem[{\citenamefont{Liebsch}(2003{\natexlab{a}})}]{Liebsch_OSMT_0}
\bibinfo{author}{\bibfnamefont{A.}~\bibnamefont{Liebsch}},
  \bibinfo{journal}{Europhysics Letters} \textbf{\bibinfo{volume}{63}},
  \bibinfo{pages}{97} (\bibinfo{year}{2003}{\natexlab{a}}).

\bibitem[{\citenamefont{Liebsch}(2004)}]{Liebsch_OSMT_2}
\bibinfo{author}{\bibfnamefont{A.}~\bibnamefont{Liebsch}},
  \bibinfo{journal}{Phys. Rev. B} \textbf{\bibinfo{volume}{70}},
  \bibinfo{pages}{165103} (\bibinfo{year}{2004}).

\bibitem[{\citenamefont{Koga et~al.}(2004{\natexlab{a}})\citenamefont{Koga,
  N.Kawakami, Rice, and Sigrist}}]{Koga_OSMT}
\bibinfo{author}{\bibfnamefont{A.}~\bibnamefont{Koga}},
  \bibinfo{author}{\bibnamefont{N.Kawakami}},
  \bibinfo{author}{\bibfnamefont{T.}~\bibnamefont{Rice}}, \bibnamefont{and}
  \bibinfo{author}{\bibfnamefont{M.}~\bibnamefont{Sigrist}},
  \bibinfo{journal}{Phys. Rev. Lett.} \textbf{\bibinfo{volume}{92}},
  \bibinfo{pages}{216402} (\bibinfo{year}{2004}{\natexlab{a}}).

\bibitem[{\citenamefont{Liebsch}(2003{\natexlab{b}})}]{Liebsch_OSMT}
\bibinfo{author}{\bibfnamefont{A.}~\bibnamefont{Liebsch}},
  \bibinfo{journal}{Phys. Rev. Lett.} \textbf{\bibinfo{volume}{91}},
  \bibinfo{pages}{226401} (\bibinfo{year}{2003}{\natexlab{b}}).

\bibitem[{\citenamefont{Koga et~al.}(2004{\natexlab{b}})\citenamefont{Koga,
  N.Kawakami, Rice, and Sigrist}}]{Koga_OSMT_SCES}
\bibinfo{author}{\bibfnamefont{A.}~\bibnamefont{Koga}},
  \bibinfo{author}{\bibnamefont{N.Kawakami}},
  \bibinfo{author}{\bibfnamefont{T.}~\bibnamefont{Rice}}, \bibnamefont{and}
  \bibinfo{author}{\bibfnamefont{M.}~\bibnamefont{Sigrist}}
  (\bibinfo{year}{2004}{\natexlab{b}}), \bibinfo{note}{preprint
  cond-mat/0406457}.

\bibitem[{\citenamefont{Castellani et~al.}(1978)\citenamefont{Castellani,
  Natoli, and Ranninger}}]{Castellani_V2O3}
\bibinfo{author}{\bibfnamefont{C.}~\bibnamefont{Castellani}},
  \bibinfo{author}{\bibfnamefont{C.~R.} \bibnamefont{Natoli}},
  \bibnamefont{and}
  \bibinfo{author}{\bibfnamefont{J.}~\bibnamefont{Ranninger}},
  \bibinfo{journal}{Phys. Rev. B} \textbf{\bibinfo{volume}{18}},
  \bibinfo{pages}{4945} (\bibinfo{year}{1978}).

\bibitem[{\citenamefont{Fr{\'e}sard and
  Kotliar}(1997)}]{fresard_multiorbital_prb_1997}
\bibinfo{author}{\bibfnamefont{R.}~\bibnamefont{Fr{\'e}sard}} \bibnamefont{and}
  \bibinfo{author}{\bibfnamefont{G.}~\bibnamefont{Kotliar}},
  \bibinfo{journal}{Phys. Rev. B} \textbf{\bibinfo{volume}{56}},
  \bibinfo{pages}{12909} (\bibinfo{year}{1997}).

\bibitem[{\citenamefont{Kotliar and Ruckenstein}(1986)}]{kotliar_ruckenstein}
\bibinfo{author}{\bibfnamefont{G.}~\bibnamefont{Kotliar}} \bibnamefont{and}
  \bibinfo{author}{\bibfnamefont{A.}~\bibnamefont{Ruckenstein}},
  \bibinfo{journal}{Phys. Rev. Lett.} \textbf{\bibinfo{volume}{57}},
  \bibinfo{pages}{1362} (\bibinfo{year}{1986}).

\bibitem[{\citenamefont{B{\"u}nemann et~al.}(2003)\citenamefont{B{\"u}nemann,
  Gebhard, and Thul}}]{gebhard_gutzwiller}
\bibinfo{author}{\bibfnamefont{J.}~\bibnamefont{B{\"u}nemann}},
  \bibinfo{author}{\bibfnamefont{F.}~\bibnamefont{Gebhard}}, \bibnamefont{and}
  \bibinfo{author}{\bibfnamefont{R.}~\bibnamefont{Thul}},
  \bibinfo{journal}{Phys. Rev. B} \textbf{\bibinfo{volume}{67}},
  \bibinfo{pages}{075103} (\bibinfo{year}{2003}).

\bibitem[{\citenamefont{{Florens} and
  {Georges}}(2002)}]{florens_rotors_imp_2002_prb}
\bibinfo{author}{\bibfnamefont{S.}~\bibnamefont{{Florens}}} \bibnamefont{and}
  \bibinfo{author}{\bibfnamefont{A.}~\bibnamefont{{Georges}}},
  \bibinfo{journal}{Phys. Rev. B} \textbf{\bibinfo{volume}{66}},
  \bibinfo{pages}{165111} (\bibinfo{year}{2002}).

\bibitem[{\citenamefont{{Florens} and
  {Georges}}(2004)}]{florens_rotor_long_2004}
\bibinfo{author}{\bibfnamefont{S.}~\bibnamefont{{Florens}}} \bibnamefont{and}
  \bibinfo{author}{\bibfnamefont{A.}~\bibnamefont{{Georges}}},
  \bibinfo{journal}{Phys. Rev. B}
  \textbf{\bibinfo{volume}{70}}(\bibinfo{number}{3}), \bibinfo{pages}{035114}
  (\bibinfo{year}{2004}).

\bibitem[{\citenamefont{{Florens} et~al.}(2002)\citenamefont{{Florens},
  {Georges}, {Kotliar}, and {Parcollet}}}]{florens_orbital_2002_prb}
\bibinfo{author}{\bibfnamefont{S.}~\bibnamefont{{Florens}}},
  \bibinfo{author}{\bibfnamefont{A.}~\bibnamefont{{Georges}}},
  \bibinfo{author}{\bibfnamefont{G.}~\bibnamefont{{Kotliar}}},
  \bibnamefont{and}
  \bibinfo{author}{\bibfnamefont{O.}~\bibnamefont{{Parcollet}}},
  \bibinfo{journal}{Phys. Rev. B} \textbf{\bibinfo{volume}{66}},
  \bibinfo{pages}{205102} (\bibinfo{year}{2002}).

\bibitem[{\citenamefont{{Florens} et~al.}(2003)\citenamefont{{Florens}, {San
  Jos{\' e}}, {Guinea}, and {Georges}}}]{florens_qdot_prb_2003}
\bibinfo{author}{\bibfnamefont{S.}~\bibnamefont{{Florens}}},
  \bibinfo{author}{\bibfnamefont{P.}~\bibnamefont{{San Jos{\' e}}}},
  \bibinfo{author}{\bibfnamefont{F.}~\bibnamefont{{Guinea}}}, \bibnamefont{and}
  \bibinfo{author}{\bibfnamefont{A.}~\bibnamefont{{Georges}}},
  \bibinfo{journal}{Phys. Rev.B}
  \textbf{\bibinfo{volume}{68}}(\bibinfo{number}{24}), \bibinfo{pages}{245311}
  (\bibinfo{year}{2003}).

\bibitem[{\citenamefont{{Georges} and {Kotliar}}(1992)}]{georges_kotliar_dmft}
\bibinfo{author}{\bibfnamefont{A.}~\bibnamefont{{Georges}}} \bibnamefont{and}
  \bibinfo{author}{\bibfnamefont{G.}~\bibnamefont{{Kotliar}}},
  \bibinfo{journal}{Phys. Rev. B} \textbf{\bibinfo{volume}{45}},
  \bibinfo{pages}{6479} (\bibinfo{year}{1992}).

\bibitem[{\citenamefont{Caffarel and Krauth}(1994)}]{caffarel_ed_prl_1994}
\bibinfo{author}{\bibfnamefont{M.}~\bibnamefont{Caffarel}} \bibnamefont{and}
  \bibinfo{author}{\bibfnamefont{W.}~\bibnamefont{Krauth}},
  \bibinfo{journal}{Phys. Rev. Lett.} \textbf{\bibinfo{volume}{72}},
  \bibinfo{pages}{1545} (\bibinfo{year}{1994}).

\bibitem[{\citenamefont{Hirsch and Fye}(1986)}]{hirsch_fye}
\bibinfo{author}{\bibfnamefont{J.~E.} \bibnamefont{Hirsch}} \bibnamefont{and}
  \bibinfo{author}{\bibfnamefont{R.~M.} \bibnamefont{Fye}},
  \bibinfo{journal}{Phys. Rev. Lett.} \textbf{\bibinfo{volume}{25}},
  \bibinfo{pages}{2521} (\bibinfo{year}{1986}).

\bibitem[{\citenamefont{de'~Medici et~al.}(2005)\citenamefont{de'~Medici,
  Georges, Kotliar, and Biermann}}]{demedici_mottpam_2005}
\bibinfo{author}{\bibfnamefont{L.}~\bibnamefont{de'~Medici}},
  \bibinfo{author}{\bibfnamefont{A.}~\bibnamefont{Georges}},
  \bibinfo{author}{\bibfnamefont{G.}~\bibnamefont{Kotliar}}, \bibnamefont{and}
  \bibinfo{author}{\bibfnamefont{S.}~\bibnamefont{Biermann}}
  \bibinfo{journal}{Phys. Rev. Lett.} \textbf{\bibinfo{volume}{95}},
  \bibinfo{pages}{066402} (\bibinfo{year}{2005}).

\bibitem{balicas_amro_2005} L.~Balicas {\it et al.},
{\it eprint cond-mat/0507457}

\bibitem{koga_0503651} A.~Koga {\it et al.}, {\sl eprint cond-mat/0503651}

\bibitem{schork_prb_1997} T.~Schork and S.~Blawid, {\sl Phys. Rev. B}
{\bf 56}, 6559, (1997).

\bibitem{sato_jpsj_2004} R.~Sato {\it et al.}, {\sl J. Phys. Soc. Jpn.}
{\bf 73}, 1864 (2004)

\bibitem{2impKondo} B. A. Jones, C. M. Varma, and J. W. Wilkins, {\sl Phys. Rev.
Lett.} 61, 125 (1988); B. A. Jones and C. M. Varma, {\sl Phys.Rev. B} 40, 324 (1989);
I. Affleck and A. W. W. Ludwig, Phys. Rev. Lett. 68, 1046
(1992); I. Affleck, A. W. W. Ludwig, and B. A. Jones,
Phys. Rev. B 52, 9528 (1995).

\bibitem{2impslave} B. A. Jones, G. Kotliar, and A. J. Millis, Phys. Rev. B
39, R3415 (1989).

\end{thebibliography}

\end{document}